\documentclass[%
 reprint,
amsmath,amssymb,
aps,
prc,
showpacs,
longbibliography]{revtex4-1}

\usepackage{graphicx}
\usepackage{epsfig} 
\usepackage{dcolumn}
\usepackage{bm}

\begin{document}

\title{Formation and decay of resonance state in $^{9}$Be and $^{9}$B nuclei.\\
Microscopic three-cluster model investigations.
}
\author{V. S. Vasilevsky}
\email{vsvasilevsky@gmail.com}
\affiliation{Bogolyubov Institute for Theoretical Physics,\\
 Kiev, Ukraine
}
\author{K. Kat\=o}
\email{kato-iku@gd6.so-net.ne.jp}
\affiliation{
 Reaction Data Centre, Faculty of Science, Hokkaido University,\\ 
Sapporo, Japan 
}
\author{N. Zh. Takibayev}
\email{takibayev@gmail.com}

\affiliation{
 Al-Farabi Kazakh National University, \\
Almaty, Kazakhstan
}



\date{\today }

\begin{abstract}
We study nature of the low-lying resonance states in mirror nuclei $^{9}$Be
and $^{9}$B. Investigations are performed within a three-cluster
model. The model makes use  of the hyperspherical harmonics, which provides
convenient description of three-cluster continuum. The dominant three-cluster 
configurations $\alpha+\alpha+n$ and $\alpha+\alpha+p$ in $^{9}$Be and $^{9}$B, 
respectively, are taken into account. Dominant decay channels for all resonance 
states in $^{9}$Be
and $^{9}$B are explored. Much attention is paid to the
controversial  $1/2^{+}$ resonance states in both nuclei. We study effects of 
the Coulomb interaction on energy and width of three-cluster resonances in the 
mirror nuclei $^{9}$Be
and $^{9}$B. We also search for the Hoyle-analogue state which is a key step 
for alternative way of $^{9}$Be and $^{9}$B syntheses in a triple collision of 
clusters in a stellar environment.
\end{abstract}

\maketitle

\section{Introduction}

The resonance state is one of the challenging problems for theoretical and 
nuclear
physics. There are common features of resonance states, observed in a few- or
many-channel systems. \ However, there are some specific features connected
with the ways of excitation or generation of resonance states \ and also in
different ways resonance state decays in nuclear systems. Special attention is
attracted by resonance states formed by three interacting clusters, i.e.
resonance states embedded in three-cluster continuum. Such resonance states
are repeatedly observed in nuclei with well-determined three-cluster
structure. These nuclei have a dominant three-cluster configuration, and it
means that bound states and many resonance states are lying bellow and above,
respectively, threshold of three-cluster continuum. In other words, bound
states and large part of resonance states in three-cluster nuclei are
generated by an interaction of three clusters. As \ examples of such nuclei,
we can mention $^{5}$H, $^{6}$He and $^{6}$Be, $^{9}$Be and $^{9}$B and many 
others.

In the present paper, a microscopic three-cluster model is applied to study 
nature
of resonance states in $^{9}$Be and $^{9}$B. Dominant three-cluster
configurations $\alpha+\alpha+n$ and $\alpha+\alpha+p$, respectively, are
selected to describe the low excitation energy region in these nuclei. The
microscopic model, which was formulated in Ref. \cite{2001PhRvC..63c4606V}, 
makes
use of the total basis of oscillator functions to describe inter-cluster
motion. The model is called as AM\ HHB which stands for the Algebraic
three-cluster Model with the Hyperspherical Harmonics Basis. The
first application of this model to study resonance structure of $^{9}$Be and
$^{9}$B was made in Ref. \cite{2014PAN..77.555N}. Results presented in Ref.
\cite{2014PAN..77.555N} were obtained with the Minnesota potential (MP). In
present paper we make use of the modified Hasegawa-Nagata potential (MHNP) 
\cite{potMHN1, potMHN2}, and we
pay much more attention to the 1/2$^{+}$ resonance states, the Coulomb effects
on resonance states in mirror nuclei. Besides, we look for the Hoyle analogue
states in $^{9}$Be and $^{9}$B.

There are many attempts to study the resonance structure of $^{9}$Be and
$^{9}$B within different methods and models see, for instance,
\cite{1989PhRvC..39.1557D,1994PAN....57.1890V,
1996PhRvC..54..132A,2001EPJA...12..413D,
2003PhRvC..68a4310A,2004NuPhA.738..342A,
1999EPJA....4...33E,2015PhRvC..92a4322O,
2016PhRvC..93e4605K}. Mainly, these investigations are performed within
the cluster model or different variants of the Resonating Group Method. In
some cases, determination of resonance parameters is carried out in the 
framework
of models when three-cluster problem is reduced to the many-channel two-body
system by representing $^{9}$Be ($^{9}$B) as coupled-channel system $^{8}%
$Be$+n$ ($^{8}$Be$+p$) and $^{5}$He$+^{4}$He ($^{5}$Li$+^{4}$He). Other group
of papers take into account that all resonance states in $^{9}$Be and $^{9}$B
belong three-cluster continuum. Position of resonance states and their
properties are determined by using the Complex Scaling Methodology or the
Hyperspherical Harmonics basis. The latter allows one to incorporate proper
boundary conditions for decay of a compound system on three independent
clusters, while the former allows one to locate resonance states in continuum
of many-channel and many-cluster systems.

Special attention is attracted by the 1/2$^{+}$ excited states in $^{9}$Be and
$^{9}$B. This is stipulated by two factors. First, position of these resonance
was obtained at different energies in various experiments. Some experiments
claimed that there are no such resonances in $^{9}$Be or 
$^{9}$B. \ Second, different theoretical investigations suggested different
energy and nature of the 1/2$^{+}$ excited states in $^{9}$Be and $^{9}$B.
Some theoretical investigations stressed there is not the 1/2$^{+}$ excited
states in $^{9}$Be and that resonance peak in reactions of the 
photodisintegration
is associated with a virtual state. Other group of investigations detected the
1/2$^{+}$ excited states in $^{9}$Be and $^{9}$B as resonance states in two-
or three-body continuum. This dispute also encouraged us to perform the
present investigations.

We need to mention numerous experimental investigations of $^{9}$Be and $^{9}%
$B: \cite{2001PhRvC..64d1305A,2010PhRvC..82a5808B,
2004PhRvC..70e4312S,2001PhRvC..63a8801U,
2000AuJPh..53..247B,1991PhRvC..43.1740G,
2014PhRvC..89b7301E,1995PhRvC..52.1315T} where structure \ and
different processes taking place in these nuclei are investigated.

Our paper is organized in the following way. In Section \ref{Sect:Model} we
explain key elements of our model. Main results are presented in Section
\ref{Sect:Results}. There is a detail discussion on nature of $1/2^{+}$
resonance states in Section \ref{Sect:12PResonan}. In Section
\ref{Sect:Coulomb} we discuss effects of the Coulomb interaction on energy and
width of resonance states in $^{9}$Be and $^{9}$B, and quest for the
Hoyle-analog states is presented in Section \ref{Sect:Hoyle}. We close the 
paper by summarizing the obtained results in Section \ref{Sect:Concl}.

\section{Model formulation}

\label{Sect:Model}

In this section we shortly outline main ideas of the model.

We start with a wave function of a nucleus consisting of three clusters, as a
key element of the model formulation. To describe a three-cluster system one 
has to construct a three-cluster function%
\begin{eqnarray}
\Psi_{JM_{J}}  &  =&\sum_{L,S}\widehat{\mathcal{A}}\left\{ \left[  \left[  \Phi
_{1}\left(  A_{1},s_{1}\right)  \Phi_{2}\left(  A_{2},s_{2}\right)  \right]
\Phi_{3}\left(  A_{3},s_{3}\right)  \right]  \right.  _{S}\nonumber\\
&  \times & \left.  f_{L,S}^{\left(  J\right)  }\left(  \mathbf{x},\mathbf{y}%
\right)  \right\}  _{JM_{J}} ~\label{eq:001}
\end{eqnarray}
and by solving a many-body Schr\"{o}dinger equation one has to determine
inter-cluster wave function $f_{L,S}^{\left(  J\right)  }\left(  \mathbf{x}%
,\mathbf{y}\right)  $ and spectrum of bound state(s) or $S$-matrix for states
of the continuous spectrum. Jacobi vectors $\mathbf{x}$ and $\mathbf{y}$ 
determine
relative position of clusters. Wave functions $\Phi_{\alpha}\left(  A_{\alpha
},s_{\alpha}\right)  $ ($\alpha$=1, 2, 3), describing internal motion of the
cluster consisted of $A_{\alpha}$ nucleons and with the spin $s_{\alpha}$, are
assumed to be fixed, and they possess some very important features, for 
instance,
they are antisymmetric and \ translation-invariant ones. Adiabaticity,
connected with a fixed form of the wave functions $\Phi_{\alpha}\left(
A_{\alpha},s_{\alpha}\right)  $, is the main assumption of the method which is
well-known as the Resonating Group Method \cite{kn:wilderm_eng}. In fact, the
wave function (\ref{eq:001}) provides a projection operator which reduces the
many-particle problem to an effective three-body problem with nonlocal and
energy-dependent potential (see detail in Ref. \cite{kn:wilderm_eng}). For
amplitudes%
\begin{equation}
f_{L,S}^{\left(  J\right)  }\left(  \mathbf{x},\mathbf{y}\right)  =\sum
_{l_{1},l_{2}}f_{l_{1},l_{2};L,S}^{\left(  J\right)  }\left(  x,y\right)
\left\{  Y_{l_{1}}\left(  \widehat{\mathbf{x}}\right)  Y_{l_{2}}\left(
\widehat{\mathbf{y}}\right)  \right\}  _{LM_{L}} \label{eq:002}%
\end{equation}
one can deduce an infinite set of the two-dimension (with respect to variables
$x$ and $y$) integro-differential equations. This set of equations can be more
simplified. If we introduce hyperspherical coordinates%
\begin{eqnarray}
x  &  =& \rho\cos\theta,\quad y=\rho\sin\theta,\label{eq:004}\\
\Omega &  =& \left\{  \theta,\widehat{\mathbf{x}},\widehat{\mathbf{y}}\right\}
\nonumber
\end{eqnarray}
and construct a full set of orthonormal hyperspherical harmonics %
\begin{equation}
\mathcal{Y}_{K,l_{1},l_{2},LM_{L}}\left(  \Omega\right)  =\chi_{K,l_{1},l_{2}}
\left(
\theta\right)  \left\{  Y_{l_{1}}\left(  \widehat{\mathbf{x}}\right)
Y_{l_{2}}\left(  \widehat{\mathbf{y}}\right)  \right\}  _{LM_{L}}
\label{eq:005}%
\end{equation}
(see
definition of the hyperspherical harmonics, for instance, in
\cite{bookJibuti84,2001PhRvC..63c4606V}),
then the wave function (\ref{eq:001}) is represented as%
\begin{widetext} 
\begin{equation}
\Psi_{JM_{J}}    = \widehat{\mathcal{A}}\left\{  \sum_{c=\{K,l_{1},l_{2};L,S\}} 
\left[ \left[
\left[  \Phi_{1}\left(  A_{1},s_{1}\right)  \Phi_{2}\left(  A_{2}%
,s_{2}\right)  \right]  \Phi_{3}\left(  A_{3},s_{3}\right)  \right]
_{S} \mathcal{Y}%
_{K,l_{1},l_{2};L}\left(  \Omega\right)\right]_{JM_{J}} 
       ~\psi_{K,l_{1},l_{2};L,S}\left(  \rho\right)    \right\}, \label{eq:006} 
\end{equation}
\end{widetext} 
where hyperradial components $\psi_{K,l_{1},l_{2};L,S}\left(  \rho\right)  $ of
the wave function obey an infinite set of integro-differential equations. The
last step toward the simplification of numerical solutions of such a system of
equations is to expand the hyperradial amplitudes \ $\left\{  \psi
_{K,l_{1},l_{2};L,S}\left(  \rho\right)  \right\}  $\ over a basis of the 
hyperradial part of oscillator functions in the six-dimension space as%
\begin{equation}
\psi_{K,l_{1},l_{2};L,S}\left(  \rho\right)  =\sum_{n_{\rho}}C_{n_{\rho}%
K,l_{1},l_{2};L,S}\left(  b\right)  R_{n_{\rho},K}\left(  \rho,b\right)  ,
\label{eq:007}%
\end{equation}
where $R_{n_{\rho},K}\left(  \rho,b\right)  $ is an oscillator function%
\begin{align}
R_{n_{\rho},K}\left(  \rho,b\right)   &  =\left(  -1\right)  ^{n_{\rho}%
}\mathcal{N}_{n_{\rho},K}r^{K}\exp\left\{  -\frac{1}{2}r^{2}\right\}
L_{n_{\rho}}^{K+3}\left(  r^{2}\right)  ,\label{eq:008}\\
r  &  =\rho/b,\quad\mathcal{N}_{n_{\rho},K}=b^{-3}\sqrt{\frac{2\Gamma\left(
n_{\rho}+1\right)  }{\Gamma\left(  n_{\rho}+K+3\right)  }},\nonumber
\end{align}
and \ $b$ is the oscillator length.

Expansion over the oscillator basis reduces the set of integro-differential
equations to the system of linear algebraic equations for expansion
coefficients%
\begin{equation}
\sum_{\widetilde{n}_{\rho},\widetilde{c}}\left\{  \left\langle n_{\rho
},c\left\vert \widehat{H}\right\vert \widetilde{n}_{\rho},\widetilde
{c}\right\rangle -E\left\langle n_{\rho},c|\widetilde{n}_{\rho},\widetilde
{c}\right\rangle \right\}  C_{\widetilde{n}_{\rho},\widetilde{c}}=0,
\label{eq:009}%
\end{equation}
where the multiple index $c$ denotes a channel of the hyperspherical basis
$c=\left\{  K,l_{1},l_{2},L,S\right\}  $. This system is relevant to bound
states and to continuous spectrum states. To obtain spectrum of bound states,
one can use diagonalization procedure for the reduced set of the equations
(\ref{eq:009}). However, to find wave functions and elements of the scattering
$S$-matrix, one has to implement  proper boundary conditions
for expansion coefficients in Eq. (\ref{eq:009}). These conditions were 
thoroughly discussed in Ref. \cite{2001PhRvC..63c4606V}.

Completeness relations for hyperspherical harmonics \ and oscillator functions
are%
\begin{align*}
\sum_{K,l_{1},l_{2},LM}\mathcal{Y}_{K,l_{1},l_{2},LM}\left(  \Omega\right)
\mathcal{Y}_{K,l_{1},l_{2},LM}\left(  \widetilde{\Omega}\right)   &
=\delta\left(  \Omega-\widetilde{\Omega}\right)  ,\\
\sum_{n_{\rho}}R_{n_{\rho},K}\left(  \rho,b\right)  R_{n_{\rho},K}\left(
\widetilde{\rho},b\right)   &  =\delta\left(  \rho-\widetilde{\rho}\right)  ,
\end{align*}
where delta function $\delta\left(  \Omega-\widetilde{\Omega}\right)  $ stands
for product of five delta functions for each of hyperspherical angles
$\left\{  \theta,\widehat{\mathbf{x}},\widehat{\mathbf{y}}\right\}  =\left\{
\theta,\theta_{\mathbf{x}},\varphi_{\mathbf{x}},\theta_{\mathbf{y}}%
,\varphi_{\mathbf{y}}\right\}  $. The completeness relations insure us that up
to now we made no restrictions or approximations. Approximations will be
formulated later, when we proceed to numerical solutions of the system of
equations (\ref{eq:009}).

Note that hyperspherical angles determine the shape and orientation in the
space of \ a triangle connecting the centers of mass of interacting clusters.
And thus, the hyperspherical harmonics describe all possible rotations and all
possible deformations of the triangle. Each hyperspherical harmonic
$\mathcal{Y}_{K,l_{1},l_{2},LM}\left(  \Omega\right)  $ (similar to the solid
harmonics) predetermines one or several dominant shapes of the three-cluster
triangles (see some illustrations for this statement in Ref.
\ \cite{2001PhRvC..63c4607V,2001PhRvC..63f4604V}).

As for functions $\psi_{K,l_{1},l_{2};L,S}\left(  \rho\right)  $ and
$R_{n_{\rho},K}\left(  \rho,b\right)  $, they describe radial excitations or
monopole or breathing mode excitations. Besides, the wave functions
$\psi_{K,l_{1},l_{2};L,S}\left(  \rho\right)  $ describes all elastic and
inelastic processes in three-cluster continuum and thus they contain elements
of the scattering $S$-matrix:%
\begin{equation}
\psi_{K,l_{1},l_{2};L,S}\left(  \rho\right)  \Rightarrow\delta_{c_{0},c}\psi
_{c}^{\left(  -\right)  }\left(  k\rho,\eta_{c}\right)  -S_{c_{0},c}\psi
_{c}^{\left(  +\right)  }\left(  k\rho,\eta_{c}\right)  , \label{eq:015}%
\end{equation}
where $c_{0}$ and $c$ denote incoming and present or outgoing channels, 
respectively; in
general case they consist of five quantum numbers $c=\left\{  K,l_{1}%
,l_{2},L,S\right\}  $. Six dimension incoming $\psi_{c}^{\left(  -\right)
}\left(  k\rho,\eta_{c}\right)  $ and outgoing $\psi_{c}^{\left(  +\right)
}\left(  k\rho,\eta_{c}\right)  $ waves are determined as follows%
\begin{equation}
\psi_{c}^{\left(  \pm\right)  }\left(  k\rho,\eta_{c}\right)  =W_{\pm
i\eta_{c},K+2}\left(  2ik\rho\right)  /\rho^{5/2}, \label{eq:016}%
\end{equation}
where $W_{\nu,\mu}\left(  z\right)  $ is the Whittaker function (see chapter
13 of book \cite{kn:abra}) and $\eta_{c}$ is the Sommerfeld parameter%
\[
\eta_{c}=\frac{m}{\hbar^{2}}\frac{Z_{c,c}e^{2}}{k}.
\]
The wave functions $\psi_{c}^{\left(  \pm\right)  }\left(  k\rho,\eta
_{c}\right)  $ are solutions to the differential equations%

\begin{align}
&  \left\{  -\frac{\hbar^{2}}{2m}\left[  \frac{d^{2}}{d\rho^{2}}+\frac{5}%
{\rho}\frac{d}{d\rho}-\frac{K\left(  K+3\right)  }{\rho^{2}}\right]  \right.
\label{eq:018}\\
&  +\left.  \frac{Z_{c,c}e^{2}}{\rho}-E\right\}  \psi_{c}^{\left(  \pm\right)
}\left(  k\rho,\eta_{c}\right)  =0,\nonumber
\end{align}
where $Z_{c,c}e^{2}/\rho$ represents the effective Coulomb interaction in
hyperspherical coordinates. It should be stressed that equation (\ref{eq:018})
and the Coulomb interaction in this equation represent an asymptotic form of
the microscopic three-cluster hamiltonian when distances between interacting
clusters are large. The effective charge $Z_{c,c}$ was determined in Ref.
\cite{2001PhRvC..63c4606V} and there was suggested an algorithm for \ its
calculation. One can find in Refs. \cite{2001PhRvC..63c4607V} and
\cite{2012PhRvC..85c4318V} the explicit values of the effective charge for
$^{6}Be$ and $^{12}C$, respectively.

The asymptotic of wave functions presented in eq. (\ref{eq:015}) is formulated
for the so-called  3-to-3 scattering. This approximation is valid when there
are no bound states in any two-cluster subsystem. That is the case for nuclei
$^{9}$Be and $^{9}$B we are going to study.

By closing this section we consider possible values of quantum numbers of the
hyperspherical harmonics. First we consider possible values of the partial
angular orbital momenta $l_{1}$ and $l_{2}$. They determine the total parity of 
a three-cluster state $\pi=\left(  -1\right)  ^{l_{1}+l_{2}}$. This implies the
first restriction on possible values of $l_{1}$ and $l_{2}$. Next, the total
orbital momentum $L$ is a vector sum of partial angular orbital momenta
$l_{1}$ and $l_{2}$: $\mathbf{L}=\mathbf{l}_{2}+\mathbf{l}_{2}$ and thus
\[
L=\left\vert l_{1}-l_{2}\right\vert ,~\left\vert l_{1}-l_{2}\right\vert
+1,~\ldots,~l_{1}+l_{2},
\]
and we have got the second restriction. the last restriction is connected with
peculiarities of the hyperspherical harmonics. For a fixed value of the
hypermomentum $K$, sum $l_{1}+l_{2}$ can have the following values%
\[
~l_{1}+l_{2}=K_{\min},K_{\min}+2,\ldots,K,
\]
where $K_{\min}=L$ for \ the normal parity state $\pi=\left(  -1\right)  ^{L}$
and $K_{\min}=L+1$ for the abnormal parity state $\pi=\left(  -1\right)
^{L+1}$. \ Combing the first and third restrictions, we conclude that \ the
hyperspherical harmonics with even values of $K$ describes positive parity
states, while harmonics with odd values of $K$ describes only negative parity 
states.

Let us consider a partial case of positive parity states with zero value of the 
total
orbital momentum $L$. In this case the partial orbital momenta $l_{1}=l_{2}%
=$0, 1, \ldots, $K/2$, and for the selected value of the hypermomentum $K$ we 
have
got $K/2+1$ hyperspherical functions. If we take all hyperspherical harmonics
with $K$=0, 2, \ldots, $K_{\max}$, we will involve $N_{ch}=\left(  K_{\max
}+2\right)  \left(  K_{\max}+4\right)  /8$ channels in our calculations, where 
$N_{ch}$ is the number of possible channels $c$. For an arbitrary
 value of the total orbital momentum $L$, by taking into account all
hyperspherical harmonics with $K=K_{\min},K_{\min}+2,\ldots,K_{\max}$, we will
involve $N_{ch}=\left(  K_{\max}-K_{\min}+2\right)  \left(  K_{\max}-K_{\min
}+4\right)  \left(  L+1\right)  /8$ channels.

As we deal with the oscillator functions, describing three-cluster system, in
the following we will use the quantum number $N_{sh}$ which numerates
oscillator shells for a state of the compound system with the parity $\pi$ and
the total angular momentum $L$. An oscillator function $\left\vert n_{\rho
},c\right\rangle $ with a given values of $n_{\rho}$ and $K$ belongs to the
oscillator shell%
\[
N_{sh}=n_{\rho}+\left(  K-K_{\min}\right)  /2
\]
provided that $K\geq K_{\min}$. \ This condition means that when $n_\rho=0$, 
the oscillator
function with hypermomentum $K=K_{\min}$ appears on the oscillator shell
$N_{sh}=0$, and the oscillator function with $K=K_{\min}+2$ appears on the 
oscillator
shell $N_{sh}=1$ and so on. If $K=K_{\min}$, then the number of the oscillator
shell $N_{sh}$ coincides with the number of the hyperradial excitations
$n_{\rho}$. On the oscillator shells $N_{sh}\geq\left(  K_{\max}-K_{\min
}\right)  /2$ we have got a fixed number of the oscillator functions $N_{f}%
$=$\left(  K_{\max}-K_{\min}\right)  /2+1$.

\section{Spectrum of resonance states in $^{9}$Be and $^{9}$B}

\label{Sect:Results}

To perform numerical calculations, we need to fix a few parameters and select
\ the nucleon-nucleon potential. We start with selection of the
nucleon-nucleon potential. We exploit the Modified Hasegawa-Nagata potential
(MHNP) \cite{potMHN1, potMHN2} to model nucleon-nucleon interaction. This is a
semi-realistic potential constructed from the realistic nuclear force by using
the reaction matrix method, and it has been intensively used in numerous
many-cluster systems, as it provides a good description of the internal
structure of clusters and interaction between clusters as well. After the NN
potential was selected, we need to fix four input parameters: oscillator
length $b$, number of channels or number of hyperspherical harmonics and
number of hyper radial excitations.

We restrict ourselves with a finite set of the hyperspherical harmonics, which
is determined by the maximal value of the hyperspherical momentum $K_{\min}$. 
To
describe the positive parity states we use all \ hyperspherical harmonics with
\ the hypermomentum $K\leq K_{\min}=14$, the negative parity states are
described by the hyperspherical harmonics with $K\leq K_{\min}=13$. \ These
amounts of the hyperspherical harmonics account for many different scenarios
of the three-cluster decay. We also restrict ourselves with number of the
hyperradial excitation $n_{\rho}\leq100$. This allows us to rich an asymptotic
region, where all clusters are well separated and the cluster-cluster
interaction, induced by the nucleon-nucleon potential, is negligible small. In
Table \ref{Tab:NChannels} we collect information about the number of total
channels ($N_{ch}$) involved in calculations for different values of the total
angular momentum $J$ and parity $\pi$. We also indicated the number of
channels compatible with the total orbital momentum $L=J-1/2$ ($N_{ch}\left(
J_{-}\right)  $) and $L=J+1/2$ ($N_{ch}\left(  J_{+}\right)  $), naturally 
$N_{ch} = N_{ch}\left(J_{-}\right) + N_{ch}\left(J_{-}\right)  $. Note that the
same set of the hyperspherical harmonics was used in Ref. 
\cite{2014PAN..77.555N}.%

\begin{table}[tbp] \centering
\caption{Number of channels involved in calculations for different states 
$J^{\pi}$  of $^9Be$ and $^9B$. $N_{ch}(J_-)$ and $N_{ch}(J_+)$ are 
explained in the text.}%
\begin{tabular}
[c]{|l|l|l|l|l|l|l|l|l|l|}\hline
$J^{\pi}$ & 1/2$^{-}$ & 1/2$^{+}$ & 3/2$^{-}$ & 3/2$^{+}$ & 5/2$^{-}$ &
5/2$^{+}$ & 7/2$^{-}$ & 7/2$^{+}$ & 9/2$^{+}$\\\hline
$K_{\max}$ & 13 & 14 & 13 & 14 & 13 & 14 & 13 & 14 & 14\\\hline
$N_{ch}\left(  J_{-}\right)  $ & -  & 21 & 28 & 12 & 21 & 44 & 42 & 30 & 54\\
$N_{ch}\left(  J_{+}\right)  $ & 29 & 12 & 21 & 44 & 42 & 30 & 30 & 54 &
36\\\hline
$N_{ch}$ & 29 & 33 & 49 & 56 & 63 & 74 & 72 & 84 & 90\\\hline
\end{tabular}
\label{Tab:NChannels}%
\end{table}%

Since the total spin of nuclei $^{9}$Be and $^{9}$B equals $S=1/2$, then the
total orbital momentum $L$ is not a quantum number and a state with the total
angular momentum $J$ will be presented by a combination of $L=J-1/2$ and
$J=L+1/2$. One may expect that the spin-orbital forces play a noticeable role
in formation of the ground and excited states in $^{9}$Be and $^{9}$B.

We selected a tree of the Jacobi vectors where the first Jacobi vector
$\mathbf{x}$ determines distance between center of mass of two alpha
particles, while the second Jacobi vector $\mathbf{y}$ indicates a distance of
the valence nucleon to the center of mass of two alpha particles. With such a
choice of the tree, the partial orbital momentum $l_{1}$ of the relative
rotation of two alpha particles has only even values. As a consequence of that
restriction, the number of independent hyperspherical harmonics is reduced
to almost 1/2, and the parity of the compound system is totally determined
by the partial orbital momentum $l_{2}$ which associated with rotation of the
valence nucleon around $^{8}Be$ as the two-cluster $\alpha+\alpha$ subsystem.

In  the present paper, as in previous calculations \cite{2014PAN..77.555N}, the
oscillator length $b$ is selected to minimize the bound state energy of the
alpha particle, which is obtained with $b$=1.317 fm. This allows us to
describe correctly the internal structure of the alpha particle. If we take
the original form of the MHNP, we obtain the
overbounded ground state in $^{9}$Be and the bound $3/2^{-}$ state in
$^{9}$B. The latter contradicts to experiments in $^{9}$B. The similar
situation was observed for the MP. \ To avoid this unphysical
situation, we changed slightly parameters of the MHNP in order to reproduce
bound state energy of $^{9}$Be. It is necessary to recall, that modification
of the Majorana parameter affects only the odd components of the central
components of the nucleon-nucleon potential. This modification does not affect
the spin-orbital components of the MHNP, which are taken in the original form.
Within the present model, the odd components determine the interaction between
clusters and are not involved in determining the internal energy of each
cluster consisting of $s$-wave configurations. Thus, by modifying the Majorana 
parameter, we obtain the correct
value of the binding energy of $^{9}$Be. This is achieved with $m$=0.4389,
which can be compared to the original value $m$ =0.4057. With this value of
the Majorana parameter, the spectrum of resonance states in $^{9}$Be and
$^{9}$B is calculated. In Table \ref{Tab:ParamMHNP} we compare original and
new parameters of the MHNP. They are presented
in the form%
\[
V_{2S+1,2T+1}(r)=\sum_{n=1}^{3}V_{2S+1,2T+1}^{\left(  n\right)  }\exp\left\{
-\frac{r^{2}}{a_{n}^{2}}\right\}  ,
\]
where $S$ and $T$ are the spin and isospin of two nucleon system, respectively.
One can see that modification of the Majorana parameter changes only the odd
components of the intermediate-range in the MHNP.%

\begin{table}[htbp] \centering
\caption{Original and new parameters of the MHNP. Intensities of the potentials 
$V_{31}$, $V_{13}$, $V_{33}$,
$V_{31}$,  are given in MeV and range of Gaussians $a$ is in fm.}%
\begin{tabular}
[c]{|c|c|c|c|c|c|c|}\hline
Parameters & \multicolumn{3}{|c|}{Original} & \multicolumn{3}{|c|}{New}%
\\\hline
$n$ & 1 & 2 & 3 & 1 & 2 & 3\\\hline
$V_{31}^{\left(  n\right)  }$ & -6.0 & -546.0 & 1655.0 & -6.0 & -546.0 &
1655.0\\
$V_{13}^{\left(  n\right)  }$ & -5.0 & -360.0 & 1144.6 & -5.0 & -360.0 &
1144.6\\
$V_{33}^{\left(  n\right)  }$ & 1.667 & -70.0 & 161.5 & 1.667 & -33.746 &
161.5\\
$V_{11}^{\left(  n\right)  }$ & 15.0 & 50.0 & 0.0 & 15.0 & 86.254 &
0.0\\\hline
$a_{n}$ & 2.500 & 0.942 & 0.542 & 2.500 & 0.942 & 0.542\\\hline
\end{tabular}
\label{Tab:ParamMHNP}%
\end{table}%

\begin{figure}
[ptbh]
\begin{center}
\includegraphics[width=\columnwidth]%
{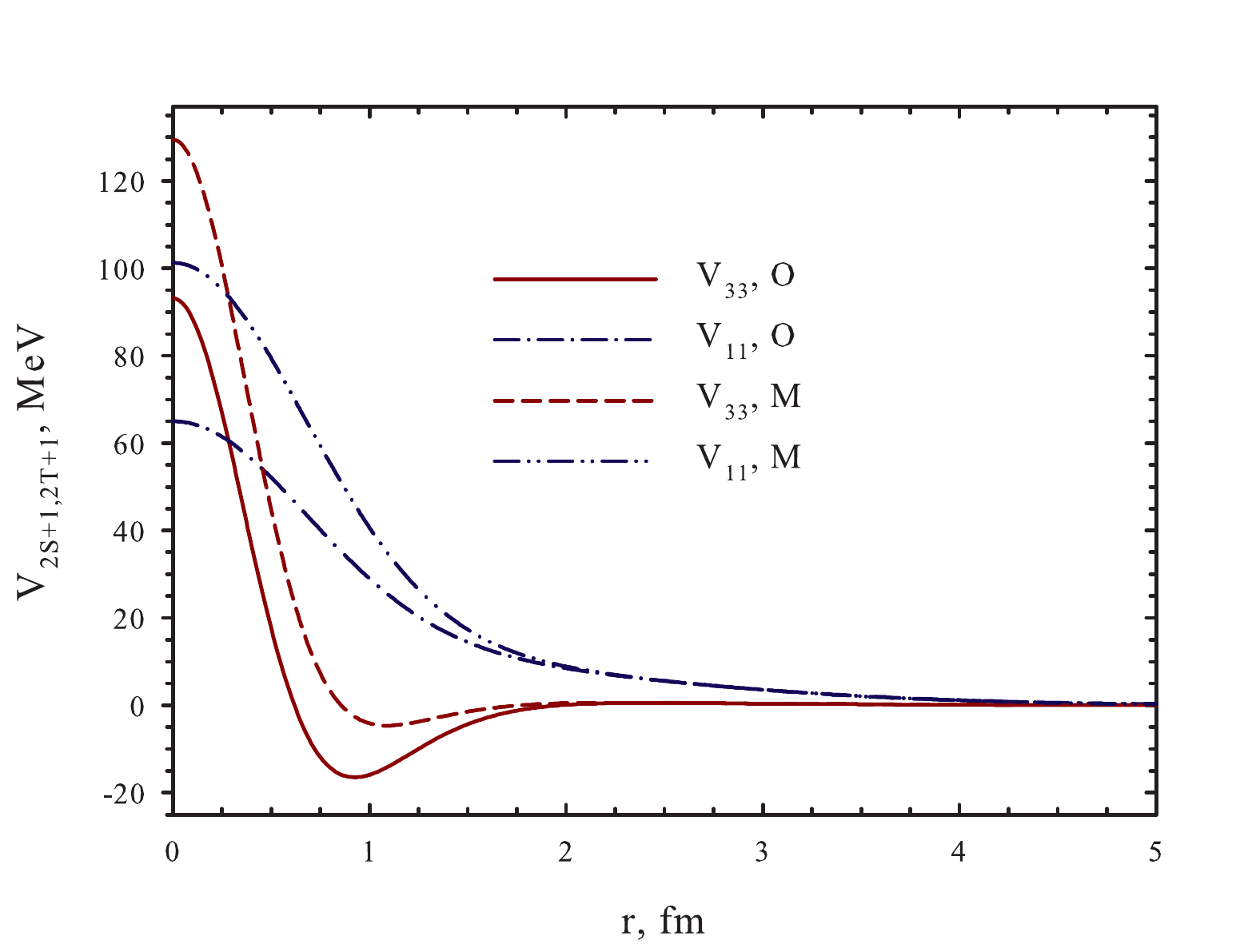}%
\caption{Odd components of the MHNP as a
function of distance between nucleons.}%
\label{Fig:MHNPoten}%
\end{center}
\end{figure}

One can also see difference between original (O) and modified (M) odd
components $V_{33}$ and $V_{11}$ of the Hasegawa-Nagata potential in Figure
\ref{Fig:MHNPoten}. Increasing of the Majorana parameter leads to a more
repulsive character of the odd components of the potential.%

\subsection{Two-cluster subsystems}

Before proceeding to resonance states in $^{9}$Be and $^{9}$B, let us consider
how the MHNP describes their two-cluster subsystems. By using new and original
parameters of the MHNP we calculate the spectrum of resonance states in
two-cluster subsystem $^{8}$Be, $^{5}$He and $^{5}$Li. \ Results of these
calculations are presented in Table \ref{Tab:2ClSpectr}. One can see that the
MHNP with original parameters describes better the spectrum of the resonance
states in $^{8}$Be, especially the $2^{+}$ and $4^{+}$ resonance states.
However, the NN potential with the original and modified sets of parameters
yield too wider $0^{+}$ resonance states with a larger energy. \ The MHNP
with the modified  Majorana parameter describes better spectrum of resonance
states in $^{5}$He and $^{5}$Li than the this potential with the original
value of $m$. However, theoretical values of energy and width of the lowest
resonance states in $^{5}$He and $^{5}$Li differ noticeable from experimental
values. Despite these discrepancies between calculated and experimental
spectrum of resonance states in two-cluster systems $^{8}$Be, $^{5}$He and
$^{5}$Li, we are going to use the MHNP with the modified value of the parameter
$m$ to study spectrum of three-cluster resonance states in compound $^{9}$Be
and $^{9}$B nuclei. As well known (see, for instance, Ref.
\cite{2014PAN..77.555N} and references therein), it is impossible to reproduce
properly spectrum of $^{9}$Be and $^{9}$B with the semi-realistic
nucleon-nucleon potential, which properly describe structure of two-cluster
subsystems $^{8}$Be, $^{5}$He and $^{5}$Li.%

\begin{table}[tbp] \centering
\caption{Spectrum of resonance states in $^8$Be, $^5$He and $^5$Li calculated 
with original (O) and modified (M) parameters of the MHNP. Energy $E$ and width 
$\Gamma$ are in MeV.}%
\begin{tabular}
[c]{|c|c|cc|cc|cc|}\hline
&  & \multicolumn{2}{|c|}{O} & \multicolumn{2}{|c|}{M} &
\multicolumn{2}{|c|}{Exp. \cite{2002NuPhA.708....3T}}\\\hline
& $J^{\pi}$ & $E$ & $\Gamma$ & $E$ & $\Gamma$ & $E$ & $\Gamma$\\\hline
$^{8}$Be & $0^{+}$ & 0.360 & 0.032 & 0.859 & 0.958 & 0.092 & 5.57$\cdot
$10$^{-6}$\\
& $2^{+}$ & 3.196 & 1.716 & 4.138 & 4.809 & 3.12 & 1.513\\
& $4^{+}$ & 11.576 & 2.569 & 14.461 & 6.386 & 11.44 & $\approx$ 3.500\\\hline
$^{5}$He & $3/2^{-}$ & -0.258 & - & 0.385 & 0.209 & 0.798 & 0.648\\
& $1/2^{-}$ & 2.307 & 10.195 & 2.335 & 11.927 & 2.068 & 5.57\\\hline
$^{5}$Li & $3/2^{-}$ & 0.608 & 0.162 & 1.236 & 0.725 & 1.69 & 1.23\\
& $1/2^{-}$ & 3.194 & 11.986 & 3.235 & 13.903 & 3.18 & 6.60\\\hline
\end{tabular}
\label{Tab:2ClSpectr}%
\end{table}%

Now we turn our attention \ to the spectrum of $^{9}$Be and $^{9}$B nuclei.
Results of \ calculations with the MHNP are presented in Tables
\ref{Tab:Resons9BeExpMHNP} and \ref{Tab:Resons9BExpMHNP} where we compare our
results with the experimental data \cite{2004NuPhA.745..155T}. Results of our
calculations are in fairly good agreement with available experimental data.
Energy and width of some resonance states are rather close to experimental
data, for instance, parameters of the 5/2$^{-}$ and 9/2$^{+}$ resonance states
in $^{9}$Be, and parameters of the 5/2$^{-}$, 1/2$^{-}$ and 5/2$^{+}$
resonance states in $^{9}$B.%

\begin{table}[htbp] \centering
\caption{Spectrum of bound and resonance states of $^9Be$ calculated with
the MHNP.}%
\begin{tabular}
[c]{|c|cc|cc|}\hline\hline
& \multicolumn{2}{|c|}{Exp.} & \multicolumn{2}{|c|}{AM\ HHB, MHNP}\\\hline
$J^{\pi}$ & $E$(MeV$\pm$keV) & $\Gamma$(MeV$\pm$keV) & $E$(MeV) & $\Gamma
$(MeV)\\\hline\hline
$3/2^{-}$ & -1.5735 &  & -1.5743 & \\
$1/2^{+}$ & 0.111$\pm7$ & 0.217$\pm$10 & 0.338 & 0.168\\
$5/2^{-}$ & 0.8559$\pm1.3$ & 0.00077$\pm0.15$ & 0.897 & 2.363$\cdot$10$^{-5}%
$\\
$1/2^{-}$ & 1.21$\pm120$ & 1.080$\pm$110 & 2.866 & 1.597\\
$5/2^{+}$ & 1.476$\pm9$ & 0.282$\pm$11 & 2.086 & 0.112\\
$3/2^{+}$ & 3.131$\pm25$ & 0.743$\pm$55 & 4.062 & 1.224\\
$3/2_{2}^{-}$ & 4.02$\pm100$ & 1.33$\pm$360 & 2.704 & 2.534\\
$7/2^{-}$ & 4.81$\pm60$ & 1.21$\pm$230 & 4.766 & 0.404\\
$9/2^{+}$ & 5.19$\pm60$ & 1.33$\pm$90 & 4.913 & 1.272\\
$5/2_{2}^{-}$ & 6.37$\pm80$ & $\sim$1.0 & 5.365 & 4.384\\
$7/2^{+}$ &  &  & 5.791 & 3.479\\\hline\hline
\end{tabular}
\label{Tab:Resons9BeExpMHNP}%
\end{table}%

%

\begin{table}[htbp] \centering
\caption{Experimental and theoretical spectrum of resonance states of $^9B$.}%
\begin{tabular}
[c]{|c|cc|cc|}\hline\hline
& \multicolumn{2}{|c|}{Exp.} & \multicolumn{2}{|c|}{AM\ HHB, MHNP}\\\hline
$J^{\pi}$ & $E$(MeV$\pm$keV) & $\Gamma$(MeV$\pm$keV) & $E$(MeV) & $\Gamma
$(MeV)\\\hline\hline
$3/2^{-}$ & 0.277 & 0.00054$\pm0.21$ & 0.379 & 1.076$\cdot$10$^{-6}$\\
$1/2^{+}$ & (1.9) & $\simeq$0.7 & 0.636 & 0.477\\
$5/2^{-}$ & 2.638$\pm5$ & 0.081$\pm$5 & 2.805 & 0.018\\
$1/2^{-}$ & 3.11 & 3.130 $\pm$ 200 & 3.398 & 3.428\\
$5/2^{+}$ & 3.065$\pm30$ & 0.550$\pm$40 & 3.670 & 0.415\\
$3/2^{+}$ &  &  & 4.367 & 3.876\\
$3/2_{2}^{-}$ &  &  & 3.420 & 3.361\\
$7/2^{-}$ & 7.25$\pm60$ & 2.0$\pm$200 & 6.779 & 0.896\\
$9/2^{+}$ &  &  & 6.503 & 2.012\\
$5/2_{2}^{-}$ & 12.670$\pm40$ & 0.45$\pm$20 & 5.697 & 5.146\\
$7/2^{+}$ &  &  & 7.100 & 4.462\\\hline\hline
\end{tabular}
\label{Tab:Resons9BExpMHNP}%
\end{table}%

\begin{figure}
[ptbh]
\begin{center}
\includegraphics[width=\columnwidth]%
{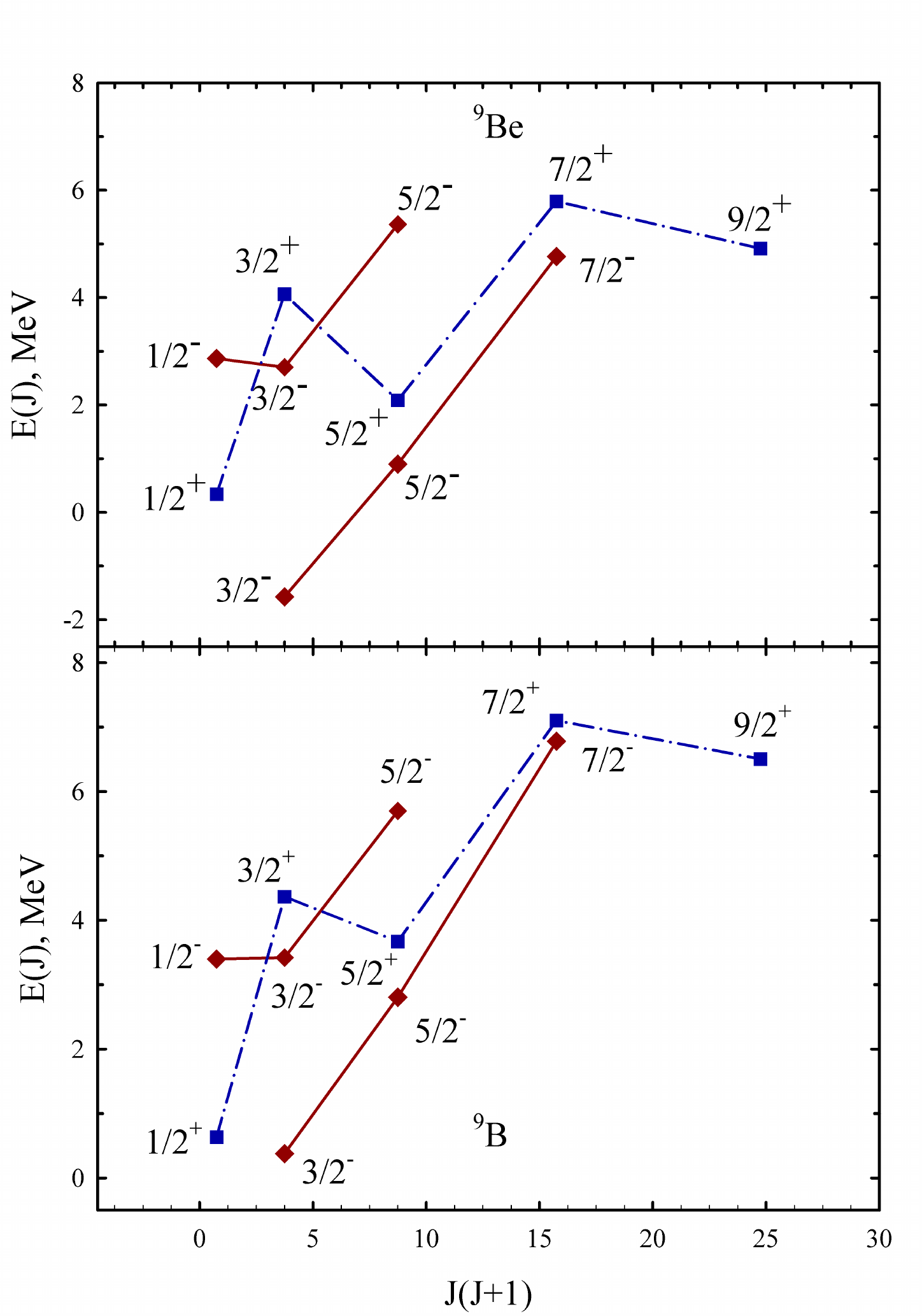}%
\caption{Spectrum of rotational bands in $^{9}$Be and $^{9}$B.}%
\label{Fig:RotatSpectr}%
\end{center}
\end{figure}

Spectrum of the ground and excited states in $^9$Be and $^9$B, presented in 
Figure \ref{Fig:RotatSpectr}, where spectrum is displayed as a function of 
$J\left(  J+1\right) $, shows that there are three rotational bands in these 
nuclei. They can be marked in the standard manner as  $K$ = 3/2$^-$, 
$K$ = 1/2$^-$, and $K$ = 1/2$^+$ rotational bands. The main $K$ = 3/2$^-$ 
rotational bands are comprised of the $3/2^{-}_1$, $5/2^{-}_1$ and $7/2^{-}_1$  
states and are represented by almost   straight lines in both nuclei. It means 
that the effective moment of inertia is of a rigid-body type, since it is 
independent of the total angular momentum $J$. This is the first rotational 
band for negative parity states. The second negative-parity rotational 
$K$ = 1/2$^-$ band involves also three states $1/2^{-}$, $3/2^{-}_2$ and 
$5/2^{-}_2$. A bent line connects these states. However, that part of 
line which connects $3/2^{-}_2$ and $5/2^{-}_2$ states, is parallel to 
the line of the first rotational band. Such shape of a line in the $K$ =1/2$^-$ 
band indicates that the Coriolis forces are strong in the 1/2$^{-}$ state 
and are very small in the 3/2$^{-}_2$ and $5/2^{-}_2$ states. The Coriolis 
forces, associated with the interaction of the internal and collective degrees 
of freedom, are rather strong in  the states of the positive parity rotational 
$K$ = 1/2$^+$ band. Especially they are very strong for the 3/2$^{+}$ and 
9/2$^{+}$ states, since these states strongly bends a line collecting states 
of this band. As in the case of the $K$ = 1/2$^-$ band, the line connecting 
the 5/2$^{+}$ and 7/2$^{+}$ is also parallel to the line of the first 
rotational 
$K$ = 3/2$^-$ band. It is interesting to note a close similarity in the 
structure of rotational bands in $^9$Be and $^9$B nuclei.

\begin{figure}
[ptbh]
\begin{center}
\includegraphics[width=\columnwidth]%
{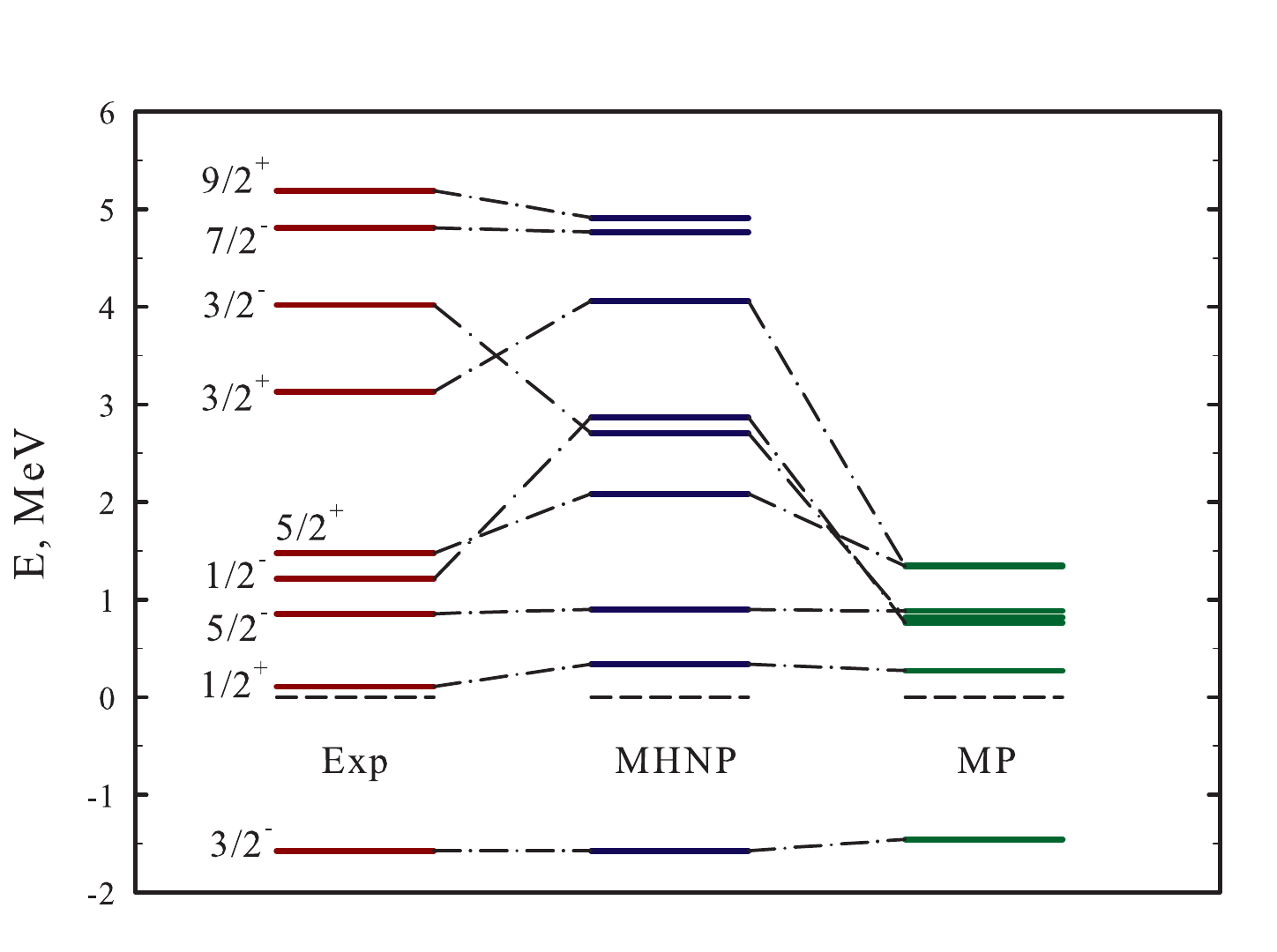}%
\caption{Experimental (Exp) and calculated spectrum of $^{9}$Be determined
with the MHNP and MP.}%
\label{Fig:Spectr9BeMHNP&MP&Exp}%
\end{center}
\end{figure}

\begin{figure}
[ptbh]
\begin{center}
\includegraphics[width=\columnwidth]%
{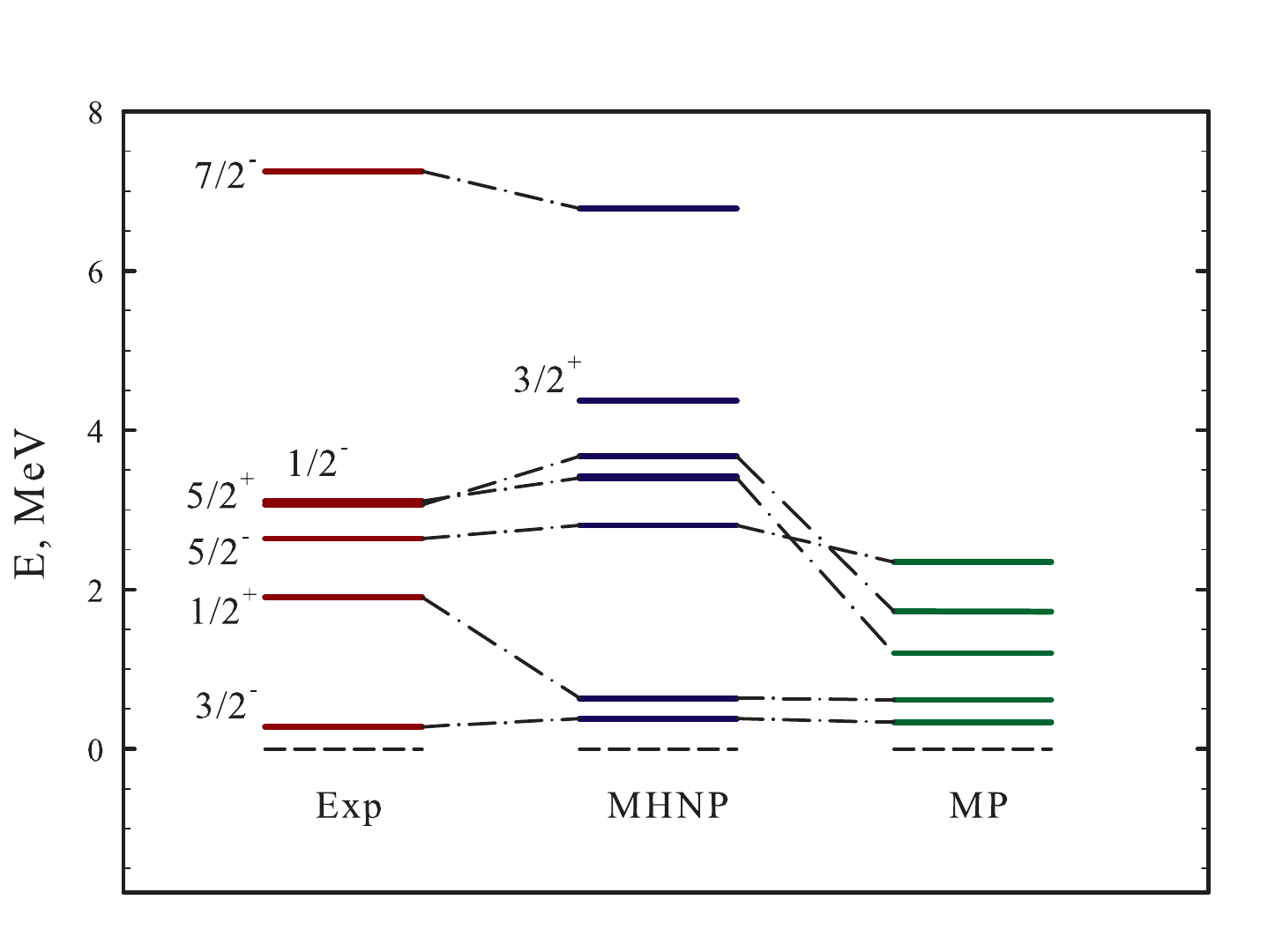}%
\caption{Experimental and calculated spectrum of $^{9}$B.}%
\label{Fig:Spectr9BMHNP&MP&Exp}%
\end{center}
\end{figure}

In Figures \ref{Fig:Spectr9BeMHNP&MP&Exp} \ and \ref{Fig:Spectr9BMHNP&MP&Exp}
we compare results of our present calculations (MHNP), with results of previous
investigations (MP), presented in \cite{2014PAN..77.555N}, and with available
experimental data \cite{2004NuPhA.745..155T}.%

There are two main differences between previous calculations with the
MP and new ones with the MHNP. First, the MHNP generates
much less (approximately two times) of resonance states within the considered
region of energy. Second, the MHNP does not create many narrow
resonance states. The common feature of these two calculations is that the
$1/2^{+}$ resonance states in $^{9}$B and $^{9}$Be are observed close to the
three-cluster threshold $\alpha+\alpha+N$. It is worth to recall, that
parameters of the MP and MHNP potentials were adjusted to reproduce energy of
the $3/2^{-}$ ground state in $^{9}$Be. With these parameters we obtained the 
$3/2^-$ resonance states of $^9$B at the energy very close to the experiment. 
This result means
that we found the correct interaction between clusters in $^{9}$B and $^{9}$Be. 
In
this paper, as in the previous one \cite{2014PAN..77.555N}, we use the same
parameters of nucleon-nucleon interactions for all other $J^{\pi}$ states.
From Figures \ref{Fig:Spectr9BeMHNP&MP&Exp} \ and
\ref{Fig:Spectr9BMHNP&MP&Exp} we conclude that the MHNP
 generates more correct cluster-cluster interaction for many sets of
the $J^{\pi}$ states, than \ the MP. \ We also conclude that
spectrum of resonance states in $^{9}$B and $^{9}$Be strongly depends on
peculiarities of the nucleon-nucleon interaction.

\begin{figure}
[ptbh]
\begin{center}
\includegraphics[width=\columnwidth]%
{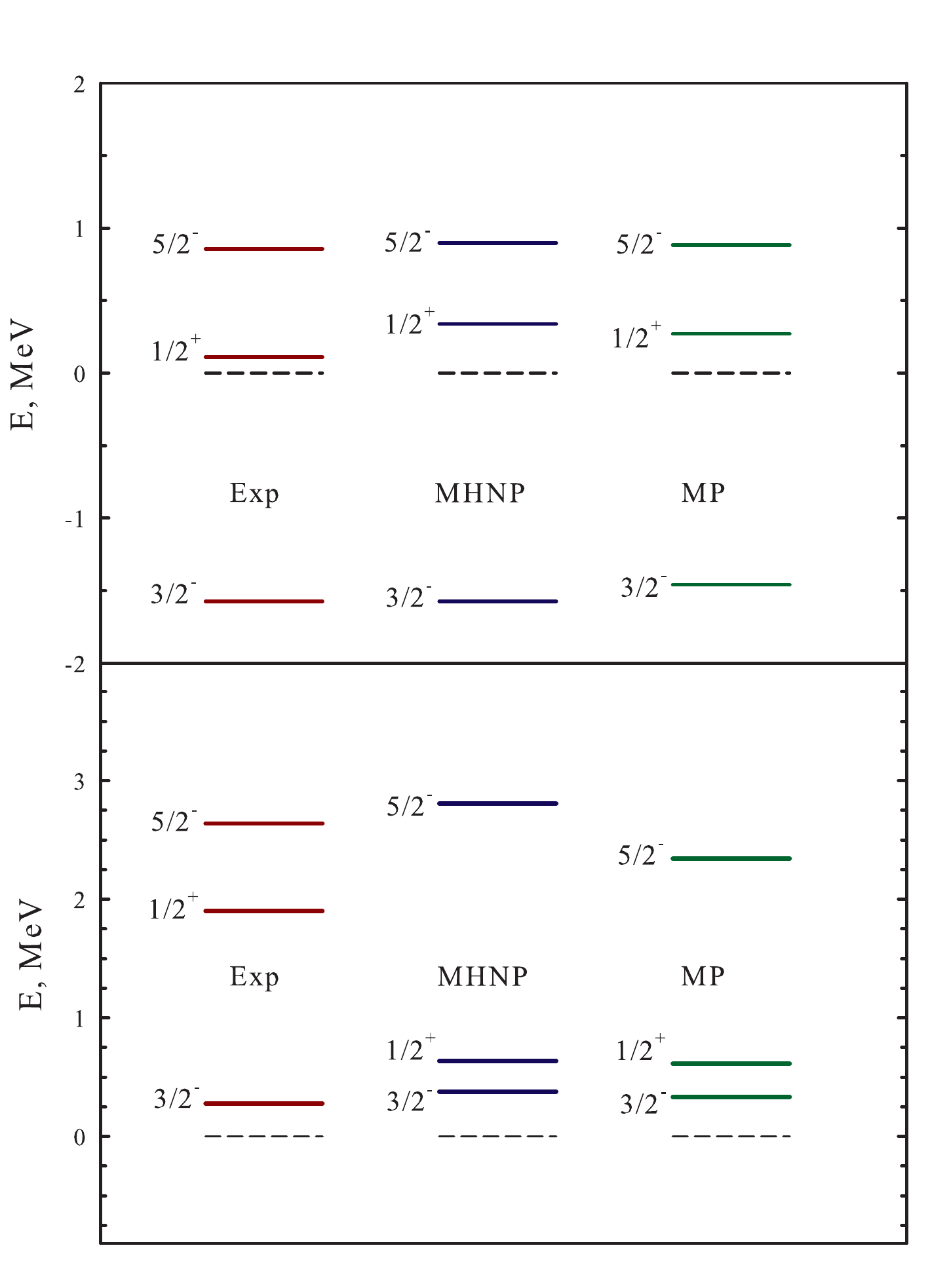}%
\caption{Comparison of some resonance states in $^{9}$Be and $^{9}$B
calculated with the MHNP and MP potentials and with experimental data.}%
\label{Fig:Spectr9Be&9BMHNPvsMPS}%
\end{center}
\end{figure}

In Figure \ref{Fig:Spectr9Be&9BMHNPvsMPS} we provide more detail comparison of
some resonance states calculated with the MHNP and MP. As we see position of
selected resonance states (namely $5/2^{-}$ and $1/2^{+}$ in both nuclei, and
$3/2^{-}$ in $^{9}$B) are almost independent or slightly dependent on the 
shape of
nucleon-nucleon potentials.%

\section{Properties of the $1/2^{+}$ resonances states}

\label{Sect:12PResonan}
\subsection{Resonance solutions in the hyperspherical harmonics method}

Now we turn our attention to the $1/2^{+}$ resonance states in $^{9}$B and
$^{9}$Be. In this section we are going to present more details about
calculations of resonance states within the present model. Numerical solutions
of the dynamic equations (\ref{eq:009}) give us $N_{ch}\times N_{ch}$ elements
of the S-matrix $\left\Vert S_{c,\widetilde{c}}\right\Vert $ \ and set of wave
functions for a given value of energy. We analyze the behavior of the diagonal
matrix elements $S_{c,c}$ of the S-matrix, which we represent as
\ $S_{c,c}=\eta_{c,c}\exp\left\{  2i\delta_{c,c}\right\}  $, where
$\delta_{c,c}$ is the phase shift and $\eta_{c,c}$\ is the inelastic
parameter. This analysis helps us to reveal resonance states and determine some
of their physical properties. However, energy of the resonance state and its 
total
width are determined by using an uncoupled channel or eigenchannel
representation which is obtained by reducing the scattering matrix $\left\Vert
S_{c,\widetilde{c}}\right\Vert $ to the diagonal form. Details of such
transformations are explained in Ref. \cite{2007JPhG...34.1955B}. The
representation allows us to calculate partial widths, to discover dominant
decay channels and thus to shed more light on nature of investigated resonance 
states.

\begin{figure}
[ptbh]
\begin{center}
\includegraphics[width=\columnwidth]%
{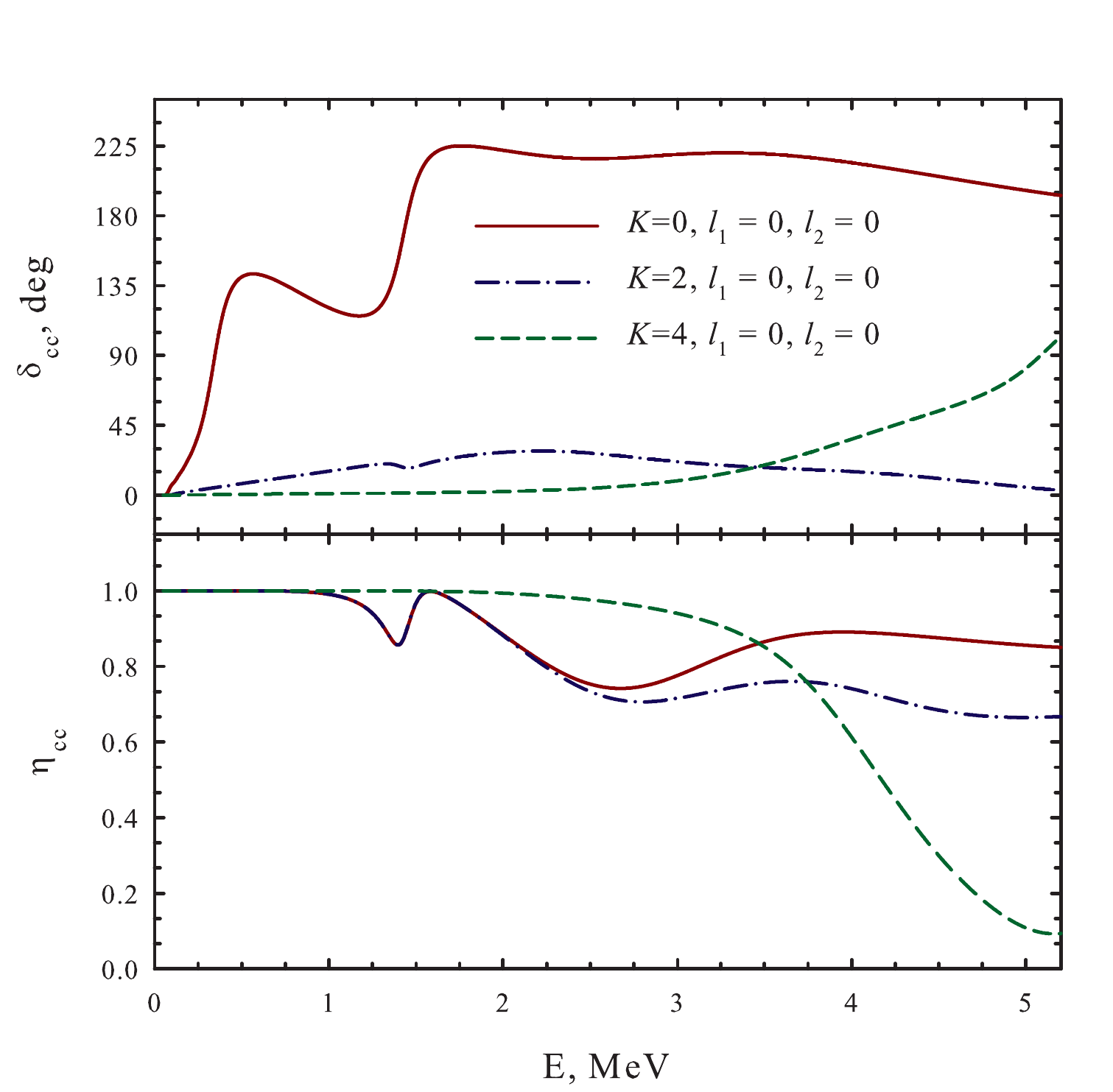}%
\caption{Phase shifts $\delta_{cc}$ and inelastic parameters $\eta_{cc}$ for
the 3$\Rightarrow$3 scattering for the $J^{\pi}=1/2^{+}$ \ state in $^{9}$Be.
}%
\label{Fig:Phases12P9BeMHNP}%
\end{center}
\end{figure}

\begin{figure}
[ptbh]
\begin{center}
\includegraphics[width=\columnwidth]%
{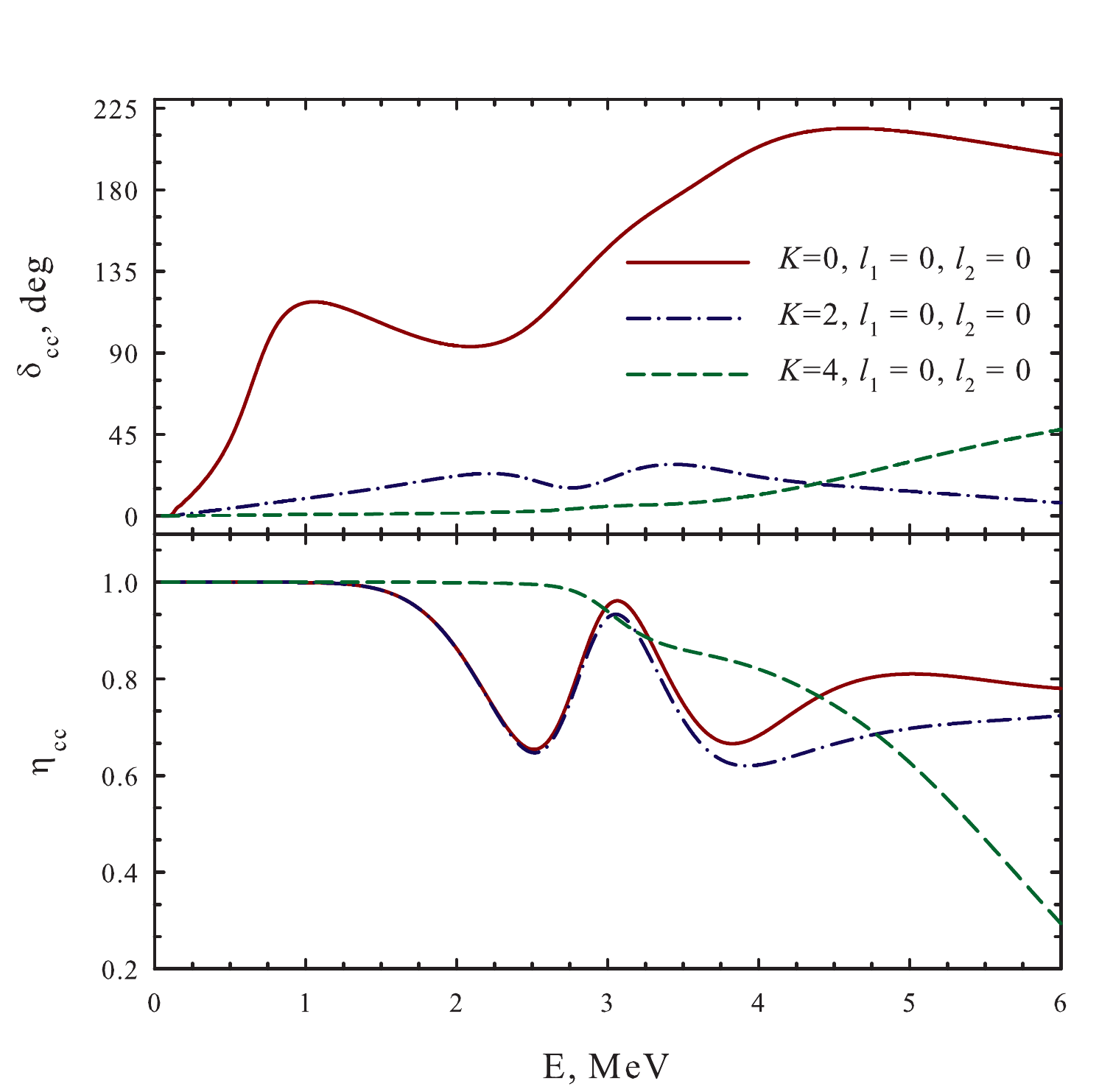}%
\caption{Phase shifts $\delta_{cc}$ and inelastic parameters $\eta_{cc}$ for
the 3$\Rightarrow$3 scattering for the $J^{\pi}=1/2^{+}$ \ state in $^{9}$B. }%
\label{Fig:Phases12P9BMHNP}%
\end{center}
\end{figure}

In Figures \ref{Fig:Phases12P9BeMHNP} and \ref{Fig:Phases12P9BMHNP} we display
the diagonal phase shifts and inelastic parameters of 3$\Rightarrow$3
scattering for the $1/2^{+}$ state in $^{9}$B and $^{9}$Be, respectively.
These results are obtained with $K_{\max}$=14 and with the MHNP. With such a
value of $K_{\max}$, 32 channels are involved in calculations (see Table
\ref{Tab:NChannels}) and only three of them produces phase shifts which are
not very small at the energy region 0$\leq E\leq$5 MeV. The phase shift 
connected
with the channel $c=\left\{  K=0,l_{1}=0,l_{2}=0,L=0\right\}  $ of $^{9}%
$Be\ shows resonance behavior at energies $E$=0.338 MeV and $E$=1.432 MeV.
The second resonance state is also reflected in the second channel $c=\left\{
K=2,l_{1}=0,l_{2}=0,L=0\right\}  $ as a shadow resonance. One should to
recall, that the shadow resonance appears in many-channel systems when it
created in one channel and reflects in other channels. The most famous shadow
resonance states are the 3/2$^{+}$ resonance states in $^{5}He$ and $^{5}Li$
which were thoroughly discussed in book \cite{kn:wilderm_eng}. \ These
resonance states are created by the Coulomb barrier in $d+t$ and $d+^{3}He$
channels, respectively, and they also reflected in $\alpha+n$ and $\alpha+p$
channels. If one disconnects $d+t$ and $\alpha+n$ or $\ d+^{3}He$ and
$\alpha+p$ channels, then one will observe the resonance state only in the
first $d+t$ or $d+^{3}He$ channel. See more detail discussion of the shadow
resonance states \ in Refs. \cite{1989PhRvC..40..902P,1993PhRvA..48.3390C}.%

Phase shifts $\delta_{cc}$ for the 1/2$^{+}$ state in $^{9}$B also exhibit
resonance states at two energies $E$=0.636 MeV and $E$=2.875 MeV. As in the
case of $^{9}$Be, resonance states in 1/2$^{+}$ state of $^{9}$B are connected
with only one channel
\[
c=\left\{  K=0,l_{1}=0,l_{2}=0,L=0\right\}  .
\]
Due to the Coulomb interaction, resonance states in $^{9}$B are shifted to a 
higher
energy region with respect to position of these resonance states in $^{9}$Be.
It is interesting to note that one observes only elastic processes around
the first $1/2^{+}$ resonance state in both nuclei, as the inelastic
parameters $\eta_{cc}=1$. Meanwhile, the elastic  and inelastic
processes are quite intensive around the second $1/2^{+}$ resonance state in
$^{9}$B and $^{9}$Be. These results mean that only one channel dominates in
formation of the first $1/2^{+}$ resonance state, and that more channels are
involved in formation and decay of the second $1/2^{+}$ resonance states.%

Phase shifts $\delta_{cc}$ for the 1/2$^{+}$ state in $^{9}$Be and $^{9}$B
demonstrate that the total orbital momentum $L=0$ dominates in this state. Can
we neglect the total orbital momentum $L$=1 to form the 1/2$^{+}$ state? What
is role of that value of the total orbital momentum? To answer these
questions, we make additional calculations by neglecting the state with $L$=1.
In this case we obtain the 1/2$^{+}$ resonance state in $^{9}$Be with
parameters $E$=0.340 MeV and $\Gamma$=0.171 MeV which is close to results with
$L$=0 and $L$=1: $E$= 0.338 MeV and $\Gamma$=0.168 MeV. In $^{9}$B we obtain
$E$=0.649 MeV and $\Gamma$=0.475 which has to be compared with $E$=0.636 MeV
and $\Gamma$=0.477. Thus, one may obtain quite correct values of energy and
total width of the 1/2$^{+}$ resonance state in $^{9}$Be and $^{9}$B by
omitting the state $L$=1.

It is important to note that with the MHNP we obtain the 1/2$^{+}$ resonance
state in $^{9}$Be and $^{9}$B with energy 0.338 and 0.636 MeV, respectively.
These energies are smaller than the energy of binary channel thresholds $^{8}%
$Be$+n$ and $^{8}$Be$+p$, if we take energy of the $0^{+}$ resonance state as
a the '"ground state" of $^{8}$Be. As we can see from Table
\ref{Tab:2ClSpectr}, this energy equals 0.859 MeV. Thus our model with the
MHNP suggests that the 1/2$^{+}$ resonance state in $^{9}$Be and $^{9}$B is of
the three-cluster nature. Let us compare what this model suggested for the MP.
Energy of the 1/2$^{+}$ resonance state in $^{9}$Be and $^{9}$B, calculated
\ in Ref. \cite{2014PAN..77.555N} with the MP, equals 0.25 and 0.59 MeV,
respectively, which is higher than the energy $E$=0.17 MeV of binary
thresholds $^{8}$Be$+n$ and $^{8}$Be$+p$. Thus the $0^{+}$ resonance state in
$^{8}$Be may participate in formation of he 1/2$^{+}$ resonance state in
$^{9}$Be and $^{9}$B.

\subsection{Resonance wave functions in the oscillator shell representation}

\subsubsection{Convergence of resonance energy and width}

To demonstrate that we involve a large set of the hyperspherical harmonics,
which provides convergent results for energy and width of resonance states, we
consider how they depend on number of hyperspherical harmonics. In Table
\ref{Tab:Converg12PResonsMHN} we demonstrated convergence of parameters of the
$1/2^{+}$ resonance state in $^{9}$B and $^{9}$Be. As we see, parameters of
the first $1/2^{+}$ resonance state in $^{9}$B and $^{9}$Be are quite stable
when we increase basis of hyperspherical harmonics from $K_{\max}=4$ to
$K_{\max}=14$. \ However, it is not the case for the second $1/2^{^{+}}$
resonance state in $^{9}$B and $^{9}$Be, as more larger set of the
hyperspherical harmonics participate in formation of the second resonance 
state.%

\begin{table}[tbp] \centering
\caption{Convergence of parameters of the $1/2^+$ resonance state in $^9Be$
and $^9B$.}%
\begin{tabular}
[c]{|l|l|l|l|l|l|l|l|}\hline
Nucleus & $K_{\max}$ & 4 & 6 & 8 & 10 & 12 & 14\\\hline\hline
$^{9}$Be & $E$, MeV & 0.338 & 0.338 & 0.338 & 0.338 & 0.338 & 0.338\\
& $\Gamma$, MeV & 0.179 & 0.175 & 0.172 & 0.171 & 0.169 & 0.168\\
& $E$, MeV & 4.972 & 3.710 & 2.091 & 1.886 & 1.764 & 1.432\\
& $\Gamma$, MeV & 3.827 & 3.869 & 1.194 & 0.641 & 0.634 & 0.233\\\hline\hline
$^{9}$B & $E$, MeV & 0.629 & 0.631 & 0.633 & 0.634 & 0.636 & 0.636\\
& $\Gamma$, MeV & 0.493 & 0.487 & 0.483 & 0.481 & 0.479 & 0.477\\
& $E$, MeV &  & 5.173 & 4.434 & 3.948 & 3.030 & 2.875\\
& $\Gamma$, MeV &  & 3.287 & 3.001 & 3.350 & 1.978 & 1.235\\\hline
\end{tabular}
\label{Tab:Converg12PResonsMHN}%
\end{table}%

Our method allows to reveal dominant decay channels by calculating partial
widths \ $\Gamma_{i}$ ($i$=1,2, \ldots, $N_{ch}$, \ $\Gamma=\sum_{i=1}%
^{N_{ch}}\Gamma_{i}$) of each resonance state. An algorithm for their
determination is presented in Ref. \cite{2007JPhG...34.1955B}. We consider
partial widths only of the second 1/2$^{+}$ resonance states in $^{9}$Be and
$^{9}$B. In Table \ref{Tab;PartWidths12P} we present total and partial widths
of those channels which make noticeable contribution to the total width of the
resonance state. In this Table we also show a ration $\Gamma_{i}/\Gamma$ in
percent ($i$= 1, 2, \ldots), which explicitly indicates the most probable decay
channels of the considered resonances. Note that the total contribution of the
presented channels exceeds 99.95 \%. 

\begin{table}[tbp] \centering
\caption{Total $\Gamma$  and partial widths $\Gamma_i$  for the second $1/2^+$ 
resonance states in $^9Be$ and $^9B$.}%
\begin{tabular}
[c]{|c|cccc|c|c|cccc|c|c|}\hline
& \multicolumn{12}{|c|}{$^{9}$Be}\\\hline
& \multicolumn{12}{|c|}{$E$= 1.432 MeV, $\ \ \Gamma$=0.2327 MeV}\\\hline
$i$ & $K$ & $l_{1}$ & $l_{2}$ & $L$ & $\Gamma_{i}$ & $\Gamma_{i}/\Gamma$ & $K$
& $l_{1}$ & $l_{2}$ & $L$ & $\Gamma_{i}$ & $\Gamma_{i}/\Gamma$\\\hline
1 & 0 & 0 & 0 & 0 & 0.1858 & 79.82 & 0 & 0 & 0 & 0 & 0.1858 & 79.82\\\hline
2 & 2 & 0 & 0 & 0 & 0.0458 & 19.67 & 2 & 0 & 0 & 0 & 0.0293 & 12.59\\\hline
3 & 4 & 2 & 2 & 0 & 0.0009 & 0.39 & 2 & 1 & 1 & 0 & 0.0165 & 7.08\\\hline
4 & 4 & 2 & 2 & 1 & 0.0001 & 0.08 & 4 & 1 & 1 & 0 & 0.0006 & 0.262\\\hline
5 &  &  &  &  &  &  & 4 & 2 & 2 & 0 & 0.0004 & 0.17\\\hline
6 &  &  &  &  &  &  & 4 & 2 & 2 & 1 & 0.0001 & 0.05\\\hline
Tree & \multicolumn{6}{|c}{$n+^{8}Be$} & \multicolumn{6}{|c|}{$^{4}He+^{5}He$%
}\\\hline\hline
& \multicolumn{12}{|c|}{$^{9}$B}\\\hline
& \multicolumn{12}{|c|}{$E$=2.871 MeV, $\ \ \Gamma$=1.2355 MeV}\\\hline
$i$ & $K$ & $l_{1}$ & $l_{2}$ & $L$ & $\Gamma_{i}$ & $\Gamma_{i}/\Gamma$ & $K$
& $l_{1}$ & $l_{2}$ & $L$ & $\Gamma_{i}$ & $\Gamma_{i}/\Gamma$\\\hline
1 & 0 & 0 & 0 & 0 & 1.1072 & 89.62 & 0 & 0 & 0 & 0 & 1.1072 & 89.62\\\hline
2 & 2 & 0 & 0 & 0 & 0.1154 & 9.34 & 2 & 0 & 0 & 0 & 0.0739 & 5.98\\\hline
3 & 4 & 0 & 0 & 0 & 0.0025 & 0.20 & 2 & 1 & 1 & 0 & 0.0416 & 3.36\\\hline
4 & 4 & 2 & 2 & 0 & 0.0088 & 0.72 & 4 & 1 & 1 & 0 & 0.0084 & 0.68\\\hline
5 & 4 & 2 & 2 & 1 & 0.0013 & 0.10 & 4 & 2 & 2 & 0 & 0.0030 & 0.24\\\hline
6 &  &  &  &  &  &  & 4 & 1 & 1 & 1 & 0.0005 & 0.04\\\hline
7 &  &  &  &  &  &  & 4 & 2 & 2 & 1 & 0.0008 & 0.07\\\hline
Tree & \multicolumn{6}{|c}{$n+^{8}Be$} & \multicolumn{6}{|c|}{$^{4}He+^{5}He$%
}\\\hline
\end{tabular}
\label{Tab;PartWidths12P}%
\end{table}%

In Tables \ref{Tab:12PResonan&Methods} and \ref{Tab:12PResonan&Methods9B} we
collect different experimental and theoretical results concerning parameters
of the 1/2$^{+}$ resonance state in $^{9}$Be and $^{9}$B. Here, ACCCM stands
for the analytic continuation in a coupling constant method
\cite{1999PhRvC..59.1391T},\ and GCM means the Generator Coordinate Method
\cite{2001EPJA...12..413D}. Both methods make use of the same part of the
Hilbert space as we use in our model. In Ref. \cite{2001EPJA...12..413D} the 
Volkov
potential N2 supplemented with the zero-range spin-orbital interaction
represents the nucleon-nucleon interaction, while the MP is
involved in the ACCCM calculations \cite{1999PhRvC..59.1391T}. The ACCCM model
generates a very broad 1/2$^{+}$ resonance state in both nuclei. Energy of
these resonances exceeds 2 MeV. Parameters of the 1/2$^{+}$ resonance states
determined within the GCM \ are close to experimental values especially for
$^{9}$Be. The GCM implies that the 1/2$^{+}$ resonance state in $^{9}$B is
broad. Within our model we obtain energy of the 1/2$^{+}$ state in $^{9}$Be is
close to experimental, however, its width is smaller than experimental. The
calculated width with the MP is very small. In $^{9}$B, our model generates
two resonance states, one of which is close to the "ground state" of the
nucleus ($E\approx$0.3 MeV) and the second one has an excitation energy $E$%
$>$ 2 MeV and a width around 1 MeV.%

\begin{table}[tbp] \centering
\caption{Parameters of the 1/2$^+$ resonance state in $^9Be$ determined by 
different experimental and theoretical methods.}%
\begin{tabular}
[c]{|c|c|c|c|}\hline
\multicolumn{4}{|c|}{$^{9}$Be}\\\hline
Method & Source & $E$, MeV & $\Gamma$, keV\\\hline
$\left(  e,e^{\prime}\right)  $ & \cite{1987ZPhyA.326..447K} & 1.684 $\pm$
0.007 & 217 $\pm$ 10\\
$\left(  e,e^{\prime}\right)  $ & \cite{1991PhRvC..43.1740G} & 1.68 $\pm$
0.015 & 200 $\pm$ 20\\
$\left(  \gamma,n\right)  $ & \cite{2001PhRvC..63a8801U} & 1.750 $\pm$ 0.010 &
283 $\pm$ 42\\\hline
$\left(  e,e^{\prime}\right)  $ & \cite{2000AuJPh..53..247B} & 1.732 &
270\\\hline
$\beta$ decay & \cite{2005NuPhA.758..647M} & 1.689 $\pm$ 0.010 & 224 $\pm$
7\\\hline
$\left(  e,e^{\prime}\right)  $ & \cite{2010PhRvC..82a5808B} & 1.748 $\pm$
0.006 & 274 $\pm$ 8\\\hline
$\left(  \gamma,n\right)  $ & \cite{PhysRevC.92.064323} & 1.728 $\pm$ 0.001 &
214 $\pm$ 7\\\hline
ACCCM & \cite{1999PhRvC..59.1391T} & 2.52 & 2620\\\hline
GCM & \cite{2001EPJA...12..413D} & 1.55 & 360\\\hline
AM HHB, MP & \cite{2014PAN..77.555N} & 1.802 & 15\\\hline
AM HHB, MHNP & Present & 1.912 & 168\\\hline
\end{tabular}
\label{Tab:12PResonan&Methods}%
\end{table}%

%

\begin{table}[tbp] \centering
\caption{Parameters of the 1/2$^+$ resonance state in $^9B$ determined by 
different experimental and theoretical methods.}%
\begin{tabular}
[c]{|c|c|c|c|}\hline
\multicolumn{4}{|c|}{$^{9}$B}\\\hline
Method & Source & $E$, MeV & $\Gamma$, MeV\\\hline
Compilation & \cite{2004PhRvC..70e4312S} & 1.0 & 1.8\\
$^{6}Li\left(  ^{6}Li,t\right)  $ & \cite{1995PhRvC..52.1315T} & 0.73 $\pm$
0.05 & $\approx$0.3\\
$^{10}B\left(  ^{3}He,\alpha\right)  $ & \cite{1988EL......5..517A} & 1.8
$\pm$ 0.2 & 0.9 $\pm$ 0.3\\\hline
$^{6}Li\left(  ^{6}Li,d\right)  $ & \cite{2012PhRvC..86c4330B} & 0.8--1.0 &
$\approx$1.5\\\hline
ACCCM & \cite{1999PhRvC..59.1391T} & 2.0 & 2.7\\\hline
GCM & \cite{2001EPJA...12..413D} & 1.27 & 1.24\\\hline
AM HHB, MP & \cite{2014PAN..77.555N} & 0.30 & 0.12\\ 
AM HHB, MP & \cite{2014PAN..77.555N} & 2.08 & 0.83\\\hline
AM HHB, MHNP & Present & 0.26 & 0.48\\
AM HHB, MHNP & Present & 2.50 & 1.24\\\hline
\end{tabular}
\label{Tab:12PResonan&Methods9B}%
\end{table}%

\subsubsection{Wave functions of resonance states}

\begin{figure}
[ptbh]
\begin{center}
\includegraphics[width=\columnwidth]%
{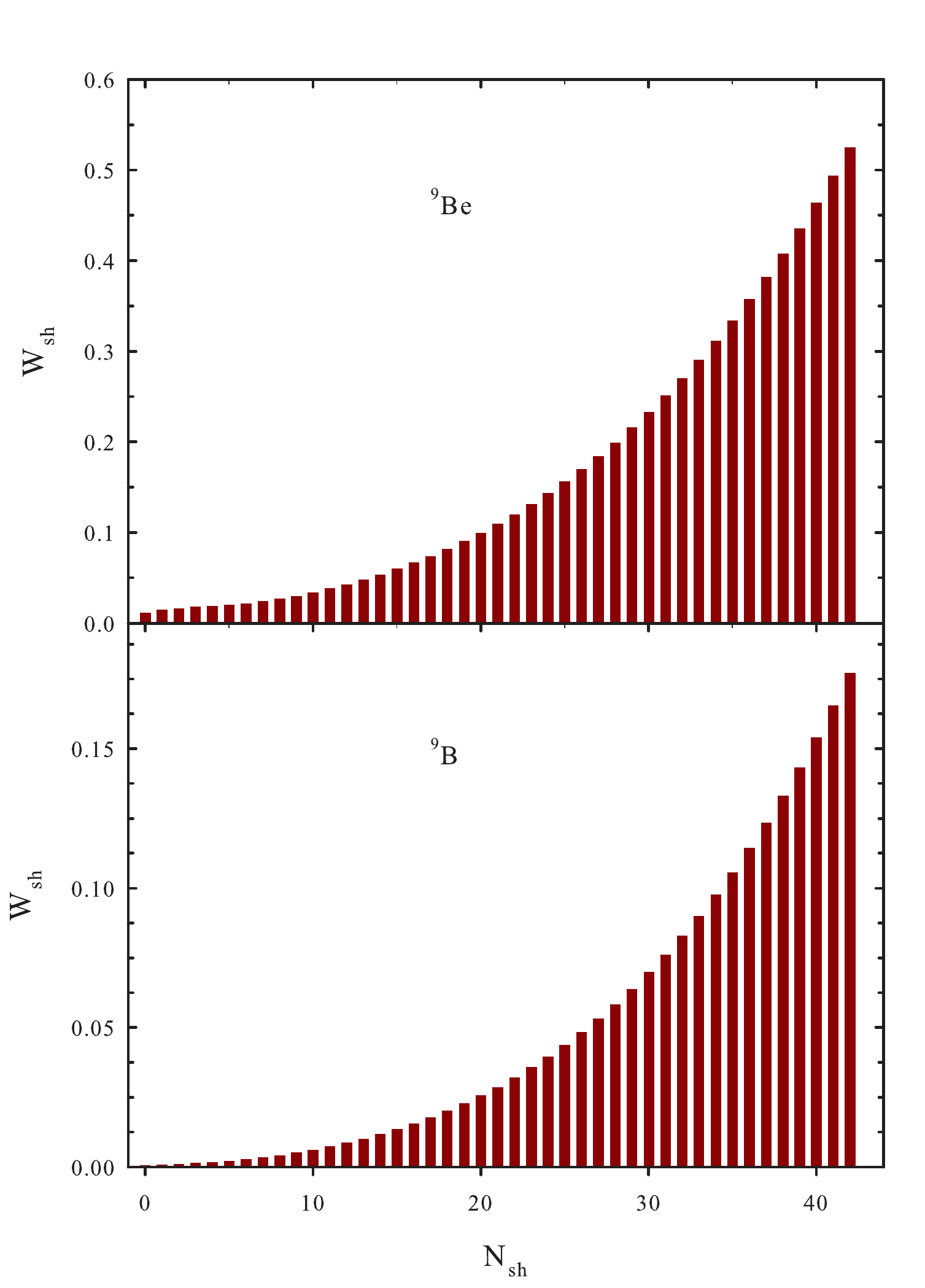}%
\caption{Weights of different oscillator shells in wave functions of 1/2$^{+}$
resonance states in $^{9}$Be and $^{9}$B.}%
\label{Fig:ShrellWeight12PRSA9MHN}%
\end{center}
\end{figure}

To understand nature of $1/2^{^{+}}$ resonance states in $^{9}$B and $^{9}$Be,
we analyze wave functions. As was mentioned above the wave function of the
three-cluster system has many-components and is a huge object which is 
difficult to analyze. The simplest way for analyzing the wave function of a 
resonance state
is to study weights of oscillator shells. In Fig.
\ref{Fig:ShrellWeight12PRSA9MHN} we show the weight $W_{sh}$ of different
oscillator shell $N_{sh}$ ($N_{sh}$ = 0, 1, 2, \ldots) in the wave function of
resonance states. The weights are determined as follows%
\[
W_{sh}=W_{sh}\left(  N_{sh}\right)  =\sum_{n_{\rho},c\in N_{sh}}\left\vert
C_{n_{\rho},c}\right\vert ^{2}.%
\]
It is important to note that oscillator wave functions with small values of
$\ N_{sh}$\ describe very compact configurations of the three-cluster system,
when distance between interacting clusters is very small. Oscillator functions
with  large values of $\ N_{sh}$\ account for configurations of \ the
three-cluster system with a large distance between all clusters and/or when one
cluster is far away from two other clusters. These statements can be deduced
from the fact that the radius (or average size) of a three-cluster system,
describing by an oscillator function from the $N_{sh}$ oscillator shell, is
equal approximately to $b\cdot\sqrt{4N_{sh}+3}$.

One can see that the wave function of the $1/2^{+}$ resonance in $^{9}$Be is
similar to the wave function of the resonance state in $^{9}$B and both of
them are represented by the oscillator shells with large values of $N_{sh}$.
Figure \ref{Fig:ShrellWeight12PRSA9MHN} displays the behavior of the wave
function which is typical for low-energy wave functions. In asymptotic region, 
which is not display here,  
these functions have an oscillatory behavior. This statement is justified by 
the following considerations. Like in a two-body case with a
sort-range interaction, the position of the first node of the wave function
shifts to at lager distances from the origin as the energy decreases to zero.
In the oscillator space we have approximately the same picture as it seen in 
the coordinate space. This is because there is a simple relation between the 
wave function in the coordinate space and expansion coefficients in the 
oscillator
representation (see detail, for instance, in \cite{2001PhRvC..63c4606V}). To
make it clear, we consider a simple case. Suppose that we involve only one
channel to describe 1/2$^{+}$ state in $^{9}$B and $^{9}$Be. This channel has
zero value of the hypermomentum and thus partial orbital momenta $l_{1}%
=l_{2}=0$ and the total orbital momentum $L$=0. An asymptotic part of the
single-channel wave function in the hyperspherical harmonic formalism is
\cite{1980CzJPh..30.1090J,1993FBS....14....1F}
\[
\psi_{K,=0}\left(  \rho\right)  \approx\frac{\cos\left(  k\rho+\delta_{0}%
-5\pi/4\right)  }{\rho^{5/2}},\quad\rho\gg1,
\]
while in the oscillator representation the expansion coefficients are
\begin{equation}
C_{n_{\rho}K=0}\left(  b\right)  \approx\sqrt{2}R_{n}^{2}\cdot\frac
{\cos\left(  kbR_{n}+\delta_{0}-5\pi/4\right)  }{R_{n}^{5/2}},\quad n_{\rho
}\gg1, \label{eq:021}%
\end{equation}
where $\delta_{0}$ is \ the phase shift of 3$\Rightarrow$3 scattering with the
hypermomentum $K$=0 and
\begin{align*}
R_{n}  &  =\sqrt{4n_{\rho}+6},\\
k  &  =\sqrt{\frac{2mE}{\hbar^{2}}}.
\end{align*}
For the sake of simplicity we assume that there is no Coulomb interaction in
both $^{9}$Be and $^{9}$B nuclei. Expansion coefficients (\ref{eq:021})
indicate that the weight of the oscillator shell $N_{sh}=n_{\rho}$ equals
approximately%
\begin{equation}
W_{n_{\rho}}\approx\left\vert \frac{\cos\left(  kbR_{n}+\delta_{0}%
-5\pi/4\right)  }{\sqrt{R_{n}}}\right\vert ^{2} \label{eq:022}%
\end{equation}
and tends to zero as $n_{\rho}$ goes to infinity due to the denominator 
proportional to $n_{\rho}^{1/2}$. It is seen from
(\ref{eq:022}) that oscillator shells which obey the conditions%
\[
kbR_{n}+\delta_{0}-5\pi/4=\nu\pi/2,\quad\nu=1,2,\ldots,
\]
give a negligibly small contribution to the wave function in the oscillator 
representation $C_{n_{\rho}K=0}$ and, consequently, to the weight 
$W_{n_{\rho}}$. One may expect a node of 
the wave function at this point of the discrete coordinate $n_{\rho}$.

In a more complicated case, when a large number of hyperspherical harmonics
are involved in calculations, one expect an oscillatory behavior of the wave
functions and weights of the oscillator shells $W_{sh}$ for any state of
continuous spectrum states. For states with large energies, more nodes of the
wave function in the oscillator representation can be observed within the
finite range of oscillator shells.%

Such a behavior of resonance wave functions may explain why these resonances
are difficult to detect by alternative methods.

It is important to note, that such shape of resonance wave function is
observed not only for the $1/2^{+}$ resonance states in $^{9}$Be and 
$^{9}$B but also for the low-lying $1/2^+$ resonances in $^{11}$B and 
$^{11}$C as shown in Ref. 
\cite{2013UkrJPh.58.544V}.
These resonance states were considered as candidates for the Hoyle
states in $^{11}$B and $^{11}$C. They are narrower ($\Gamma=$12 and $\Gamma
=$163 keV in $A=11$ \ comparing to $\Gamma=$168 and $\Gamma=$477 keV in $A=9$)
than the $1/2^{+}$ resonance states in $^{9}$Be and $^{9}$B, however, the 
behavior of shell
weights is very similar.

\begin{figure}
[ptbh]
\begin{center}
\includegraphics[width=\columnwidth]%
{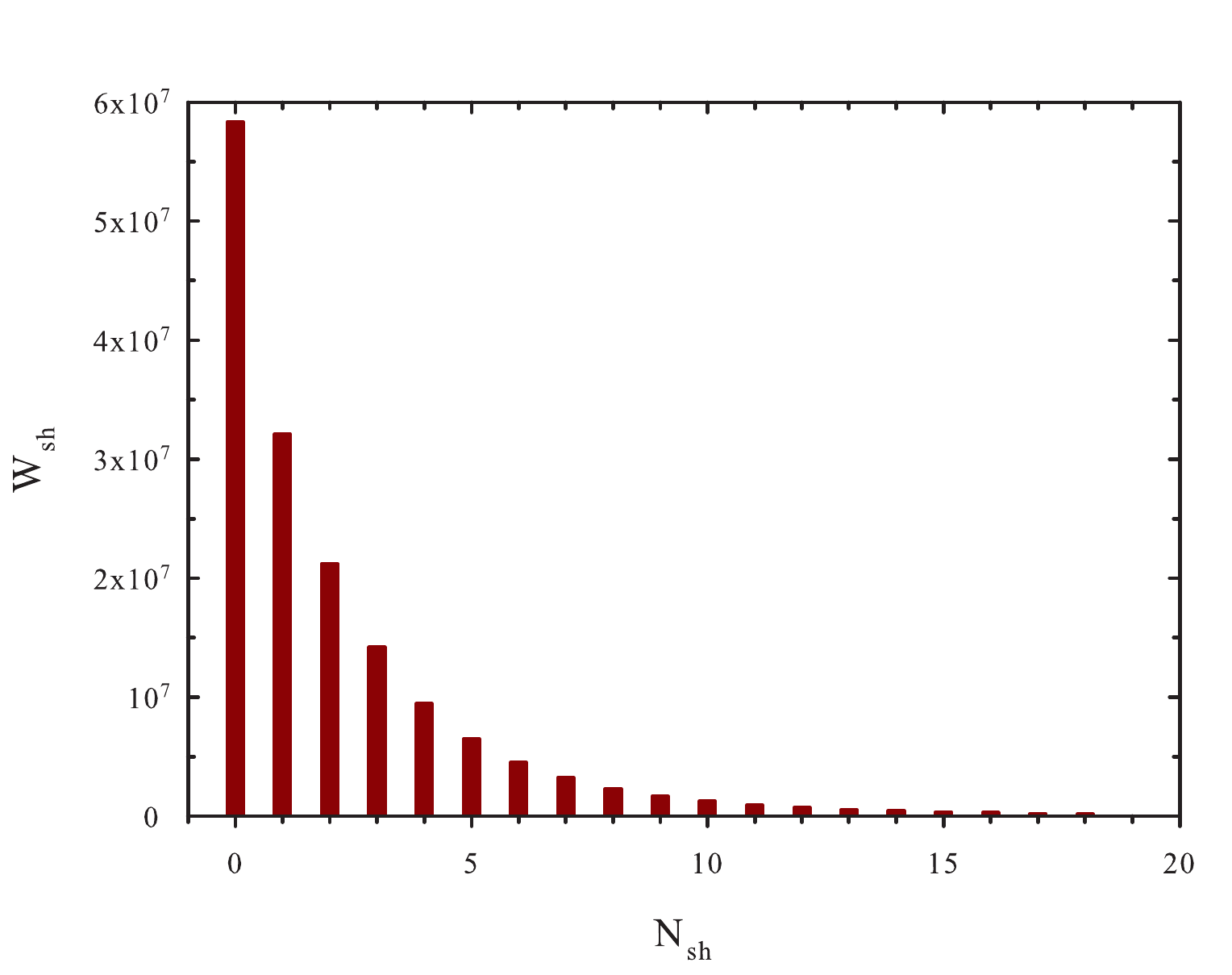}%
\caption{Structure of wave function of the $3/2^{-}$ resonance state in
$^{9}$B.}%
\label{Fig:ShrellWeight32MRSA9BHN}%
\end{center}
\end{figure}

In Fig. \ref{Fig:ShrellWeight32MRSA9BHN} we show the weights $W_{sh}$ for the
narrow $3/2^{-}$ resonance state in $^{9}$B. This is a typical picture for
very narrow resonance states in light nuclei. There are two main features of
wave functions of narrow resonance states. First, the wave function is
represented by the oscillator shells with small values of $N_{sh}$. Second, 
it has very large values of the weights $W_{sh}$ for resonance states.%

It should be stressed that all wave functions of continuous spectrum states
are normalized by the standard condition%
\[
\left\langle \Psi_{E,J^{\pi}}|\Psi_{\widetilde{E},J^{\pi}}\right\rangle
=\delta\left(  k-\widetilde{k}\right)  ,
\]
where%
\[
k=\sqrt{\frac{2mE}{\hbar^{2}}},\quad\widetilde{k}=\sqrt{\frac{2m\widetilde{E}%
}{\hbar^{2}}}.
\]

By analyzing the total and partial widths, we determine the dominant decay
channels of a three-cluster resonance state. This analysis help us to shed
some light on the nature of a resonance channel in many-channel systems. It
can be performed for two different trees of the Jacobi vectors, which were
denoted as $n+^{8}$Be and $^{4}$He$+^{5}$He in Ref. \cite{2014PAN..77.555N}.
The $1/2^{+}$ resonance state in $^{9}$Be and $^{9}$B has only dominant
channel. In the first tree, the resonance prefer to decay into the channel,
where the relative orbital momentum of two alpha particles and the orbital
momentum of valence neutron (with respect to the center of mass of two alpha
particles) equal zero. Partial widths connected with that channel almost
coincides with the total width. The same situation is observed in the second
tree. There is also only one dominant channel with zero values of partial
orbital momenta. The first orbital momentum represents relative motion of
neutron around first alpha particle and the second one represents relative
motion of the second alpha particle with respect to the center of mass of the
subsystem $\alpha+n$. These properties of the $1/2^{+}$ resonance states in
$^{9}$Be and $^{9}$B are based on two important factors. The first factor is 
the
dominant role of the channel with the hypermomentum $K=0$ in wave function of
the resonance state. The second factor is connected with the essential
properties of the hyperspherical harmonics with $K=0$. With this value of
hypermomentum, we have got only one hyperspherical harmonic which is
independent on choice of the Jacobi vector tree.

\subsection{Resonances versus attraction}

To shed some more light on structure of the $1/2^{+}$ states in $^{9}$Be and
$^{9}$B, we are going to increase attractive effective interaction in these 
nuclei by
manipulating with the exchange parameter of nucleon-nucleon forces. By
increasing the effective interaction in a many-channel system we expect to
decrease energy and width of resonance states to persuade ourselves that the
$1/2^{+}$ states in $^{9}$Be and $^{9}$B are resonance states. For virtual
states such an increase of parameters or increase of attraction between
clusters will lead to appearance of the corresponding bound states.

\begin{figure}
[ptbh]
\begin{center}
\includegraphics[width=\columnwidth]%
{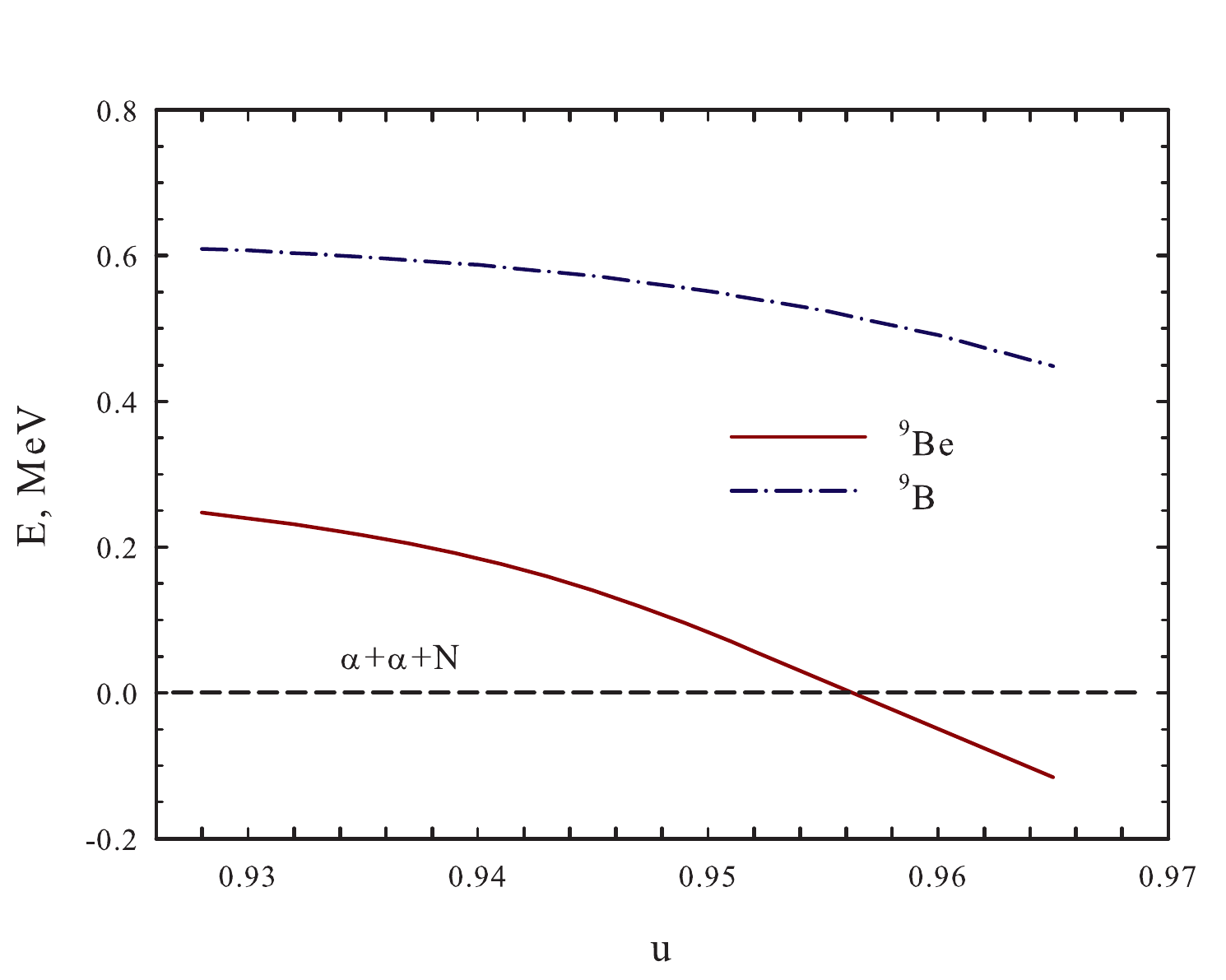}%
\caption{Energy of $1/2^{+}$ resonance state in $^{9}$Be and $^{9}$B as a
function of parameter $u$ of the MP.}%
\label{Fig:Energy12PRSvsU}%
\end{center}
\end{figure}

\begin{figure}
[ptbh]
\begin{center}
\includegraphics[width=\columnwidth]%
{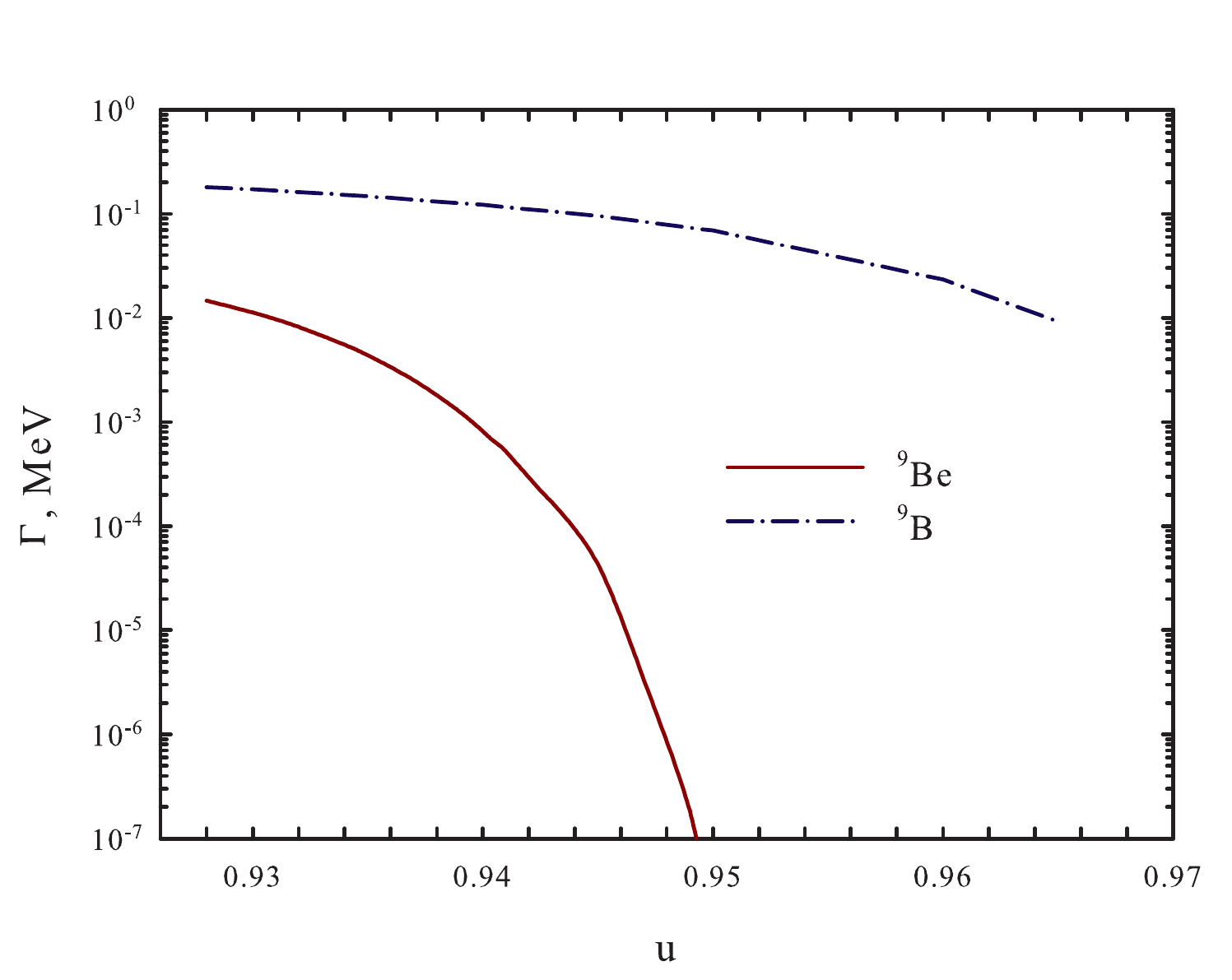}%
\caption{Dependence of the width of the $1/2^{+}$ resonance state in $^{9}$Be
and $^{9}$B on the parameter $u$ of the MP.}%
\label{Fig:Width12PRSvsU}%
\end{center}
\end{figure}

\begin{figure}
[ptbh]
\begin{center}
\includegraphics[width=\columnwidth]%
{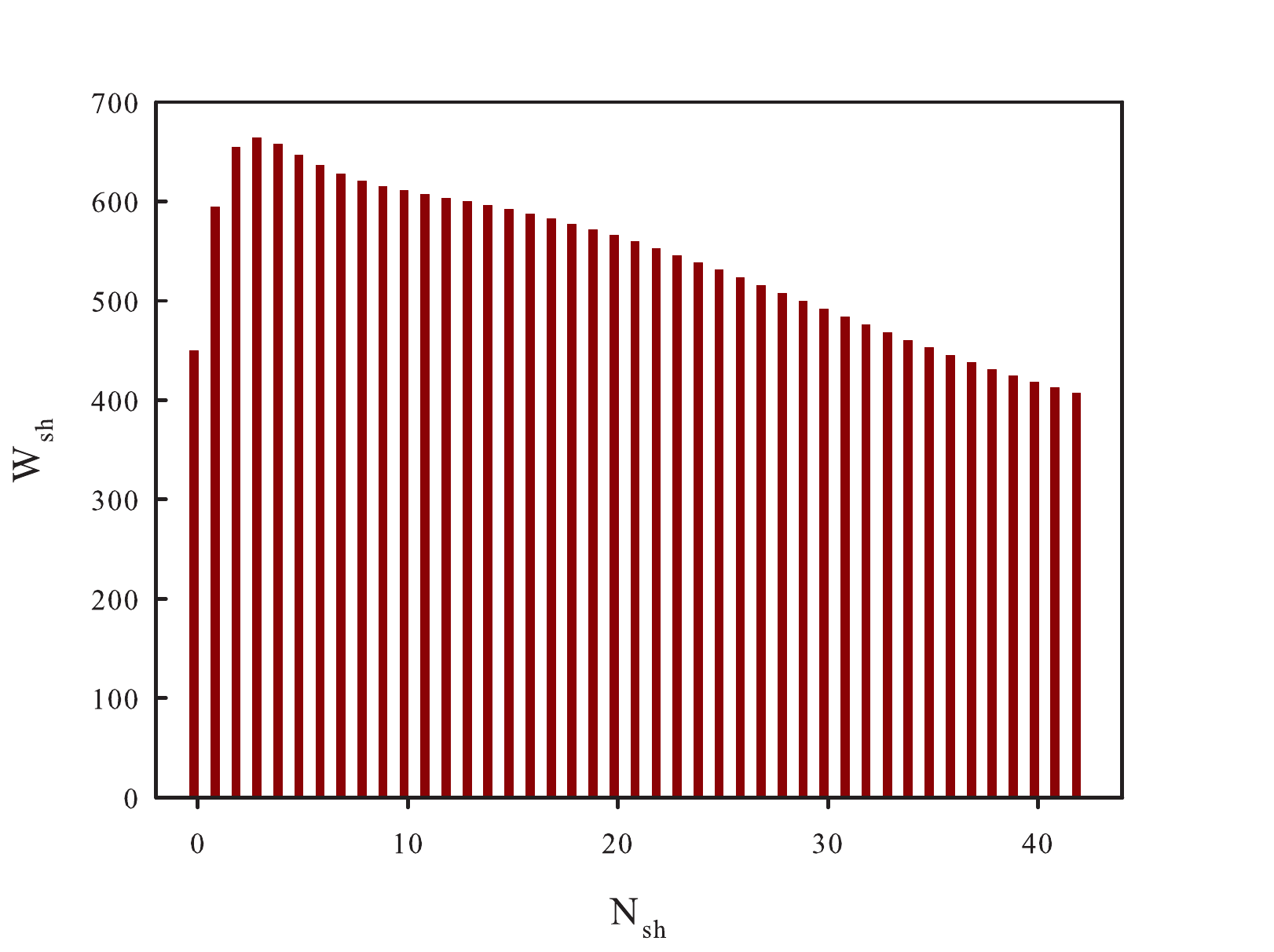}%
\caption{Structure of wave function of the $1/2^{+}$ resonance states in
$^{9}$Be calculated with the MP and $u$=0.945.}%
\label{Fig:ShellWeights12PRS9BeU0945}%
\end{center}
\end{figure}

In Ref \cite{2012PhRvC..85c4318V} such the procedure was used to study
dependence of energy and width of the $0^{+}$ resonance state in $^{12}C$
(well-known as the Hoyle state) on the exchange parameter $u$ of the MP. It was 
shown that increasing of $u$ (which results in increasing of
the effective interaction in $^{12}$C) leads to decreasing of energy and width
of the $0^{+}$ resonance states, created by the interaction between three
alpha particles. Here we use the same procedure and also the same
nucleon-nucleon potential (namely the MP). The form of this
potential allows to reduce significantly numeric calculations of parameters of
$1/2^{+}$ resonance states as a function of $u$. We also believe that the same
dependence of resonance parameters upon the exchange parameter of the
nucleon-nucleon force can be obtained for the MHNP. In Figures
\ref{Fig:Energy12PRSvsU} and \ref{Fig:Width12PRSvsU} we display how energy and
width of the $1/2^{+}$ resonance states in $^{9}$Be and $^{9}$B\ depend on the
parameter $u$. \ Note that we start our calculations with $u$=0.928, which
gives energy of the $^{9}$Be ground state $E$=-1.59 MeV, and end up with
$u$=0.945 which yields the bound state energy of $^{9}$Be $E$=-2.36 MeV. This
proves that we indeed increase the effective interaction (attraction) between
clusters in $^{9}$Be and $^{9}$B by increasing the parameter $u$ of the
MP. As we see from Figure \ref{Fig:Energy12PRSvsU} that such
a manipulation with the parameter $u$ slightly decreases energy of the
$1/2^{+}$ resonance state in $^{9}$B and more strongly reduces \ (almost two
times) energy of the resonance state in $^{9}$Be. Increasing of the effective
attraction leads to more substantial changes of widths of $1/2^{+}$ resonance
states. As we see from Figure \ref{Fig:Width12PRSvsU}\ that the width of the
$1/2^{+}$ resonance state in $^{9}$B \ is diminished from 181 keV to 95 keV,
while the width of the $1/2^{+}$ resonance state in $^{9}$Be drops from 15 keV 
to
92 eV. Thus the $1/2^{+}$ resonance state in $^{9}$Be became very a narrow
resonance state which resides close to the $\alpha+\alpha+n$ threshold.
However, the excitation energy of this state measured from the $^{9}$Be ground 
states is
increased from $E$=1.80 MeV ($u$=0.928) to $E$=2.50 MeV ($u$=0.945).  It is
interesting to note that a very narrow resonance state has a completely
different wave function comparing to the wave function with starting values of
the parameter $u$=0.928 or the wave function obtained with the MHNP. In Figure
\ref{Fig:ShellWeights12PRS9BeU0945} we display weights of different oscillator
shells in the wave function of a very narrow $1/2^{+}$ resonance state in
$^{9}$Be. Comparing this Figure \ref{Fig:ShellWeights12PRS9BeU0945} with Figure 
\ref{Fig:ShrellWeight12PRSA9MHN}, we see that the narrow resonance state has a 
large amplitude of wave
function in the internal region and its wave function is represented by
oscillator shells with small values of $N_{sh}$.%

It is important to make the following comment. Because the binding energy of 
alpha clusters does not depend on the $u$-parameter, the threshold energy of 
$\alpha+\alpha+n(p)$ is independent of $u$.
But the threshold energies of $^8$Be+$n(p)$ and $^5$He($^5$Li)+$\alpha$ depend 
on $u$. 
The $1/2^+$ states in $^9$Be and $^9$B are calculated above the three-body 
$\alpha+\alpha+n(p)$ threshold but below the two-body threshold. Therefore, 
the calculated $1/2^+$ states are unbound for the three-body 
$\alpha+\alpha+n(p)$ threshold but bound state for the two-body threshold.  
The virtual state, which was observed, for instance, in Refs. 
\cite{2015PhRvC..92a4322O,2016PhRvC..93e4605K}, is usually defined for the 
two-body threshold.

Results of the present section indicated that the $1/2^{+}$ states in $^{9}$Be
and $^{9}$B are indeed resonance states. Besides, the analysis of  dominant
decay channels and wave functions leads us to the conclusion that the
three-body effects, originated from nucleon-nucleon interaction and the
Coulomb potential between protons, and the Pauli principle plays very
important role in formation of the $1/2^{+}$ resonance states in $^{9}$Be and
$^{9}$B.

\section{Effects of Coulomb forces}
\label{Sect:Coulomb}

Within the present model, differences in the position  and
 of the total width of resonance states in mirror nuclei $^{9}$Be and $^{9}$B 
arise from the
Coulomb interaction solely, which is more stronger in $^{9}$B than in $^{9}$Be.

Effects of the Coulomb interaction on mirror or isobaric nuclei have been
repeatedly investigated by many authors. Very often influence of the Coulomb
potential on the spectrum of such nuclei is associated with the Thomas-Erhman
effect or shift (see, for instance, \cite{2007EPJA...34..315H} and references
therein), which is connected with the shift of energy of single particle
levels in mirror nuclei due to the Coulomb interaction.  By considering the
mirror nuclei in the isotopic spin formalism, one can suggest  two-fold
effects of the Coulomb forces on parameters of resonance states. First,
increasing of the Coulomb interaction leads to decreasing of the attractive 
effective
interaction in each channel of a many-channel system. That may shift up energy
of resonance states and may also increase width of resonance states. 
Second,  the Coulomb
interaction makes an effective barrier more wider, that may decrease both
energy and width of resonance state. What scenario is realized in nuclei
$^{9}$Be and $^{9}$B and how it depends on the total angular momentum $J$?

\begin{figure}
[ptbh]
\begin{center}
\includegraphics[width=\columnwidth]%
{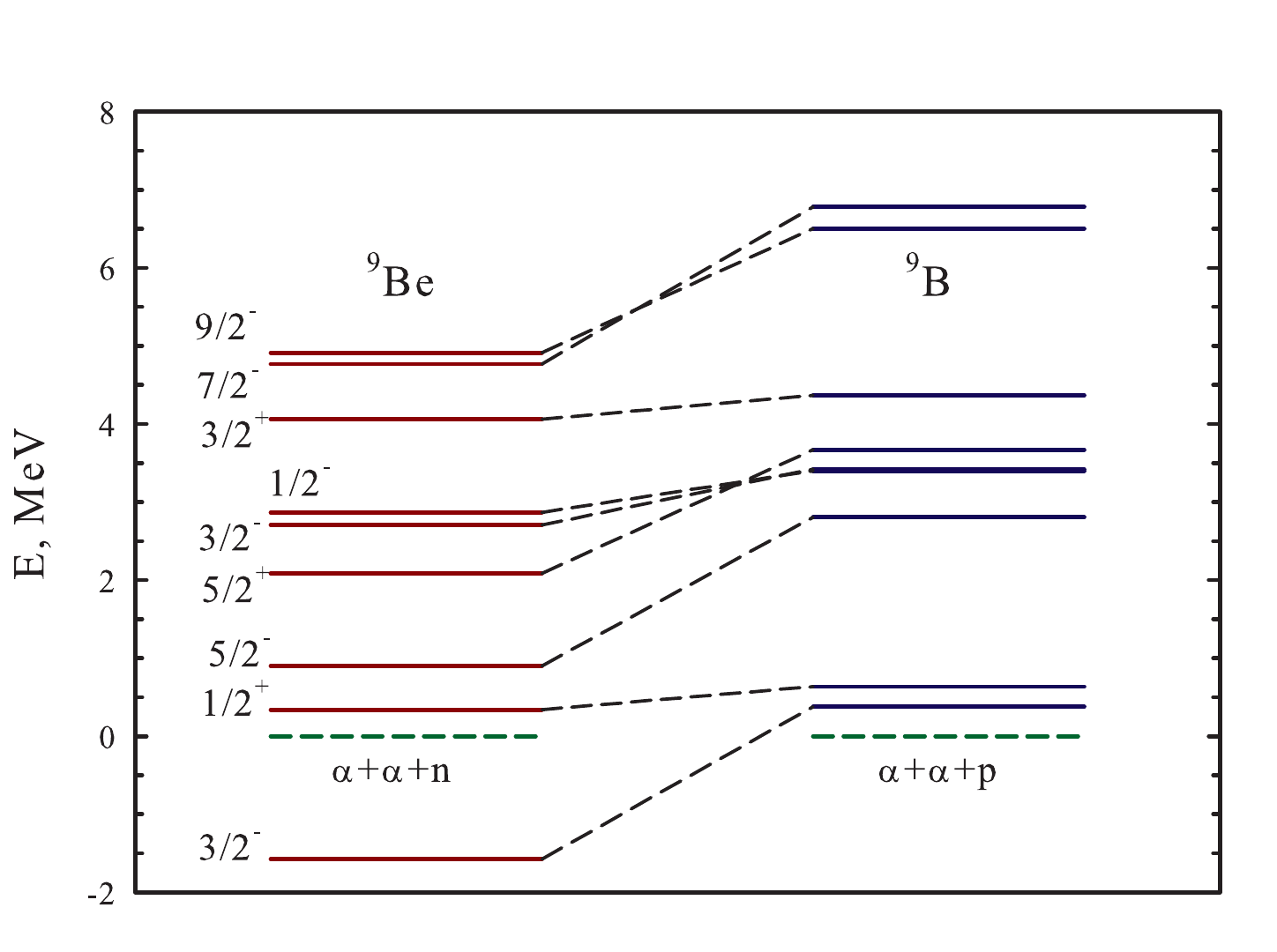}%
\caption{Spectrum of bound and resonance states in $^{9}$Be and $^{9}$B,
calculated with the MHNP.}%
\label{Fig:Spectr9Be&9BMHNP}%
\end{center}
\end{figure}

A first effects of the Coulomb forces in the mirror nuclei
$^{9}$Be and $^{9}$B \ can be seen in Figure \ref{Fig:Spectr9Be&9BMHNP}
where spectrum of these nuclei is shown. Five dashed lines, connecting levels
with the same total angular momentum $J$ and parity $\pi$ in $^{9}$Be and
$^{9}$B, show that Coulomb forces significantly shift up levels ($J^{\pi}$=
3/2$^{-}$, 5/2$^{-}$, 5/2$^{+}$, 7/2$^{-}$ and 9/2$^{-}$) and four dashed
lines indicate a moderate shift up of energy of resonance states ($J^{\pi}$=
1/2$^{+}$, 3/2$^{-}$, 1/2$^{-}$ and 3/2$^{+}$) in $^{9}$B comparing with
correspondent states in $^{9}$Be.%

\begin{figure}
[ptbh]
\begin{center}
\includegraphics[width=\columnwidth]%
{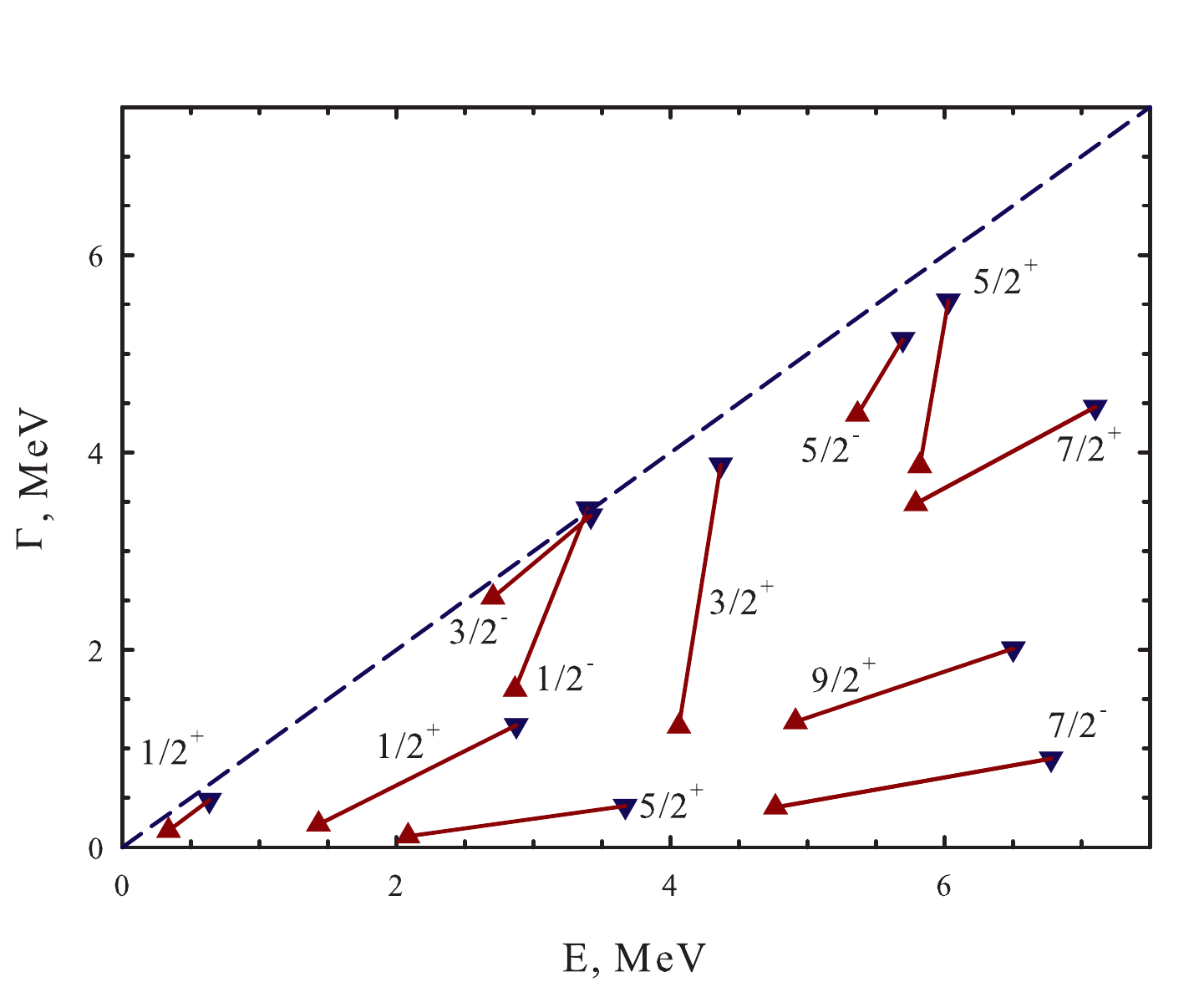}%
\caption{Displacement of resonance states due to Coulomb interaction. }%
\label{Fig:Energy&GammaMHNP}%
\end{center}
\end{figure}

In Figure \ref{Fig:Energy&GammaMHNP} we present how Coulomb forces affects
both energy\ $E$ and width $\Gamma$ of resonance states. In this Figure
resonance states of $^{9}$Be are marked by triangle up, while resonance states
of $^{9}$B in the plane $E$ -$\Gamma$ are indicated by triangle down. Dashed
line in the Figure indicates situation when the width of a resonance state 
equals to
its energy $\Gamma=E$. As one can see from Tables \ref{Tab:Resons9BeExpMHNP},
\ \ref{Tab:Resons9BExpMHNP} and from Figure \ref{Fig:Energy&GammaMHNP} that
energy of the resonance state of $^{9}$Be with given total angular momentum $J$
and parity $\pi$\ is lower  than the analogue resonance state in $^{9}%
$B. All resonance states in $^{9}$Be are more narrow than corresponding
resonance states in $^{9}$B.%

Results presented in Tables \ref{Tab:Resons9BeExpMHNP},
\ \ref{Tab:Resons9BExpMHNP} and Figures \ref{Fig:Spectr9Be&9BMHNP},
\ref{Fig:Energy&GammaMHNP} indicate that the first scenario is realized in
mirror nuclei $^{9}$Be and $^{9}$B. The Coulomb forces leads to the increasing
both energy and width of resonance states. We can split all resonance states
on three categories depending on effects of the Coulomb interaction. For this
aim we calculate a "Coulomb shift angle"
\[
\theta_{C}=\arctan\left(  \frac{E\left(  ^{9}B\right)  -E\left(
^{9}Be\right)  }{\Gamma\left(  ^{9}B\right)  -\Gamma\left(  ^{9}Be\right)
}\right)
\]
for each pair of resonance states in $^{9}$Be and $^{9}$B with the given values 
of the total momentum $J$ and parity $\pi$.  The first category consists of
resonance states with the Coulomb shift angle  $40\leq\theta_{C}\leq50$
degrees.\ We call this category as a category with  the moderate Coulomb
effects. There are several resonance states when Coulomb forces increase
energy and width in the same proportion. This leads to the case when the line
connected analogue resonance states is parallel or almost parallel to the line
$\Gamma=E$. These are 1/2$^{+}$ and 3/2$^{-}_2$ resonance states. The second
category of resonance states consists of resonance states with the Coulomb
shift angle $\theta_{C}<37$ degrees and we call it as a category of weak
Coulomb effects. In this case the Coulomb interaction changes strongly energy
and  weakly changes resonance width. This category includes   5/2$^{-}_1$,
5/2$^{+}_1$, 7/2$^{-}$, 7/2$^{+}$ and 9/2$^{+}$ resonance states. The third
category consists of resonance states with slightly changed energy but with
strongly increased width. The Coulomb shift angle for this category
$\theta_{C}>66$  degrees. This category consists of four resonance states:
1/2$^{-}$,\ 3/2$^{+}$,\ 5/2$^{-}_2$ and 5/2$^{+}_2$.

Thus, the Coulomb interaction has week, moderate or strong influence on
parameters of resonance states in mirror nuclei $^{9}$Be and $^{9}$B.

\section{Hoyle analog states}

\label{Sect:Hoyle}

We recall that the Hoyle state is a very narrow resonance state in $^{12}C$.
It lies not far from the three-cluster threshold ($E$= 0.38 MeV) and has very
small width  $\Gamma=$ 8.5 eV. One of the main features of the Hoyle
resonance state that it is a very long-lived resonance state (according to
nuclear scale). The Hoyle state is a dominant way for the nucleosynthesis of
carbon in helium-burning red giant stars, which are rich of alpha particles.
The present AM\ HHB model was successfully used in Ref.
\cite{2012PhRvC..85c4318V} to study a spectrum of bound and resonance states in
$^{12}$C. It was obtained that the Hoyle state is generated by the triple
collision of three alpha particles. (It should be stressed that the present 
model is accounted both for the sequential and simultaneous decay or excitation 
of a three-cluster resonance state. Technically it is very difficult to 
distinguish the sequential decay from simultaneous one. Thus, in our notion 
a term ``the triple
collision'' stands for both types of processes.) These results of the present 
model encouraged us to search the Hoyle-analogue states in $^{9}$Be. There is a
quest for the Hoyle-analogue states in light nuclei by different theoretical
methods. Here, we are going to find in $^{9}$Be the Hoyle-analogue state(s) by
using the AM\ HHB model. 

If we look at Table \ref{Tab:Resons9BeExpMHNP}, we find that $^{9}$Be has two
resonance states (1/2$^{+}$ and 5/2$^{-}$) which lie close to the
three-cluster threshold $\alpha+\alpha+n$. However, the 1/2$^{+}$ resonance
state is not narrow one, as ratio $\Gamma/E$ is large ($\Gamma/E\approx$ 0.5).
Meanwhile, the 5/2$^{-}$ resonance state is indeed narrow resonance state
because width is small $\Gamma$=23.6$\ $eV and besides ratio $\Gamma/E$ is
also very small: it equals $\Gamma/E\approx$ 2.63$\cdot$10$^{-5}$ in our model
and the experimental ratio is $\Gamma/E\approx$ 9.0$\cdot$10$^{-4}$. One can
compare this ratio with the experimental ratio $\Gamma/E\approx$ 2.24$\cdot
$10$^{-7}$ for the Hoyle state.

Our calculation indicates that the 5/2$^{-}$ resonance state is 
the Hoyle-analogue state.
This state has quite large half-life time, it could emit quadrupole gamma
quanta and transit to the ground state of $^{9}$Be. This is one of possible
ways for synthesis of $^{9}$Be. We assume, that in stars with large densities
of alpha particles and neutrons this is a very plausible way of creating $^{9}%
$Be nuclei.

The present model also indicates that the 1/2$^{+}$ resonance state, 
being too wide or too
short-lived one, has very small chance to participate in synthesis of $^{9}$Be.

\begin{figure}
[ptbh]
\begin{center}
\includegraphics[width=\columnwidth]%
{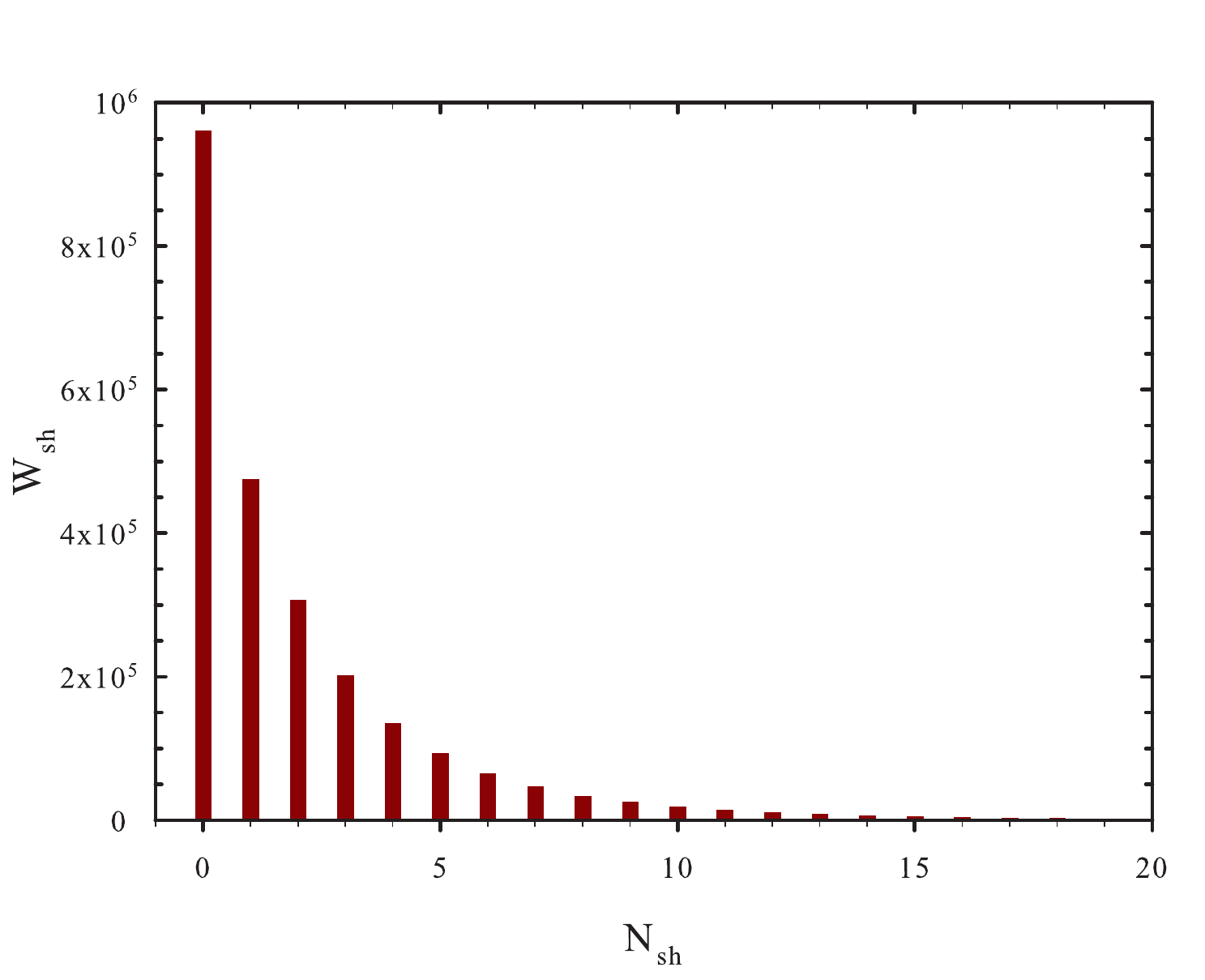}%
\caption{Weights of different oscillator shells in the wave function of the
5/2$^{-}$ resonance state in $^{9}$Be.}%
\label{Fig:ShrellWeight52PRS9BeMHN}%
\end{center}
\end{figure}

Let us consider the structure of wave function of the 5/2$^{-}$ resonance
state in $^{9}$Be. In Figure \ref{Fig:ShrellWeight52PRS9BeMHN} we demonstrate
the weight $W_{sh}$ of different shells in the wave function of the 5/2$^{-}$
resonance state. It can be concluded from the Figure that the 5/2$^{-}$
resonance state is a compact object, as it mainly represented by the
oscillator shells with small number of $N_{sh}$. The structure of wave function 
of the 5/2$^{-}$ resonance is  similar to the structure of the $^{9}$Be ground 
state wave function, which was displayed in Figure 3 in Ref. 
\cite{2014PAN..77.555N}. In both cases, the main contribution to wave functions 
of bound and resonance states comes from the oscillator shells $N_{sh} \le $15. 
Besides, the wave function of the
resonance state has a very large amplitude in the internal region ($W_{sh}$
$\leq10^{6}$). Such a behavior of the wave function of the 5/2$^{-}$ resonance 
state in $^{9}$Be is very similar to the behavior of the wave function of the 
Hoyle state in
$^{12}$C (see, for instance, Refs. \cite{2013UkrJPh.58.544V,
2008IJMPE..17.2005N,2009FBS....45..145N}).%

By comparing Figures \ref{Fig:ShrellWeight32MRSA9BHN} and
\ref{Fig:ShrellWeight52PRS9BeMHN} for resonance states 3/2$^{-}$ in $^{9}$B
and 5/2$^{-}$ in $^{9}$Be, respectively, we came to the conclusion that
\ these figures represent the standard behavior of wave function for narrow
resonance states. It means that the wave function of a ``standard'' resonance
 state has a very large amplitude in the internal region, this amplitude is 
much larger than 
the oscillating amplitude in the asymptotic regions. It also means that three
 clusters spend long time in 
the region, where inter-cluster distances are small and an interaction between 
them is strong.

\begin{figure}
[ptbh]
\begin{center}
\includegraphics[width=\columnwidth]%
{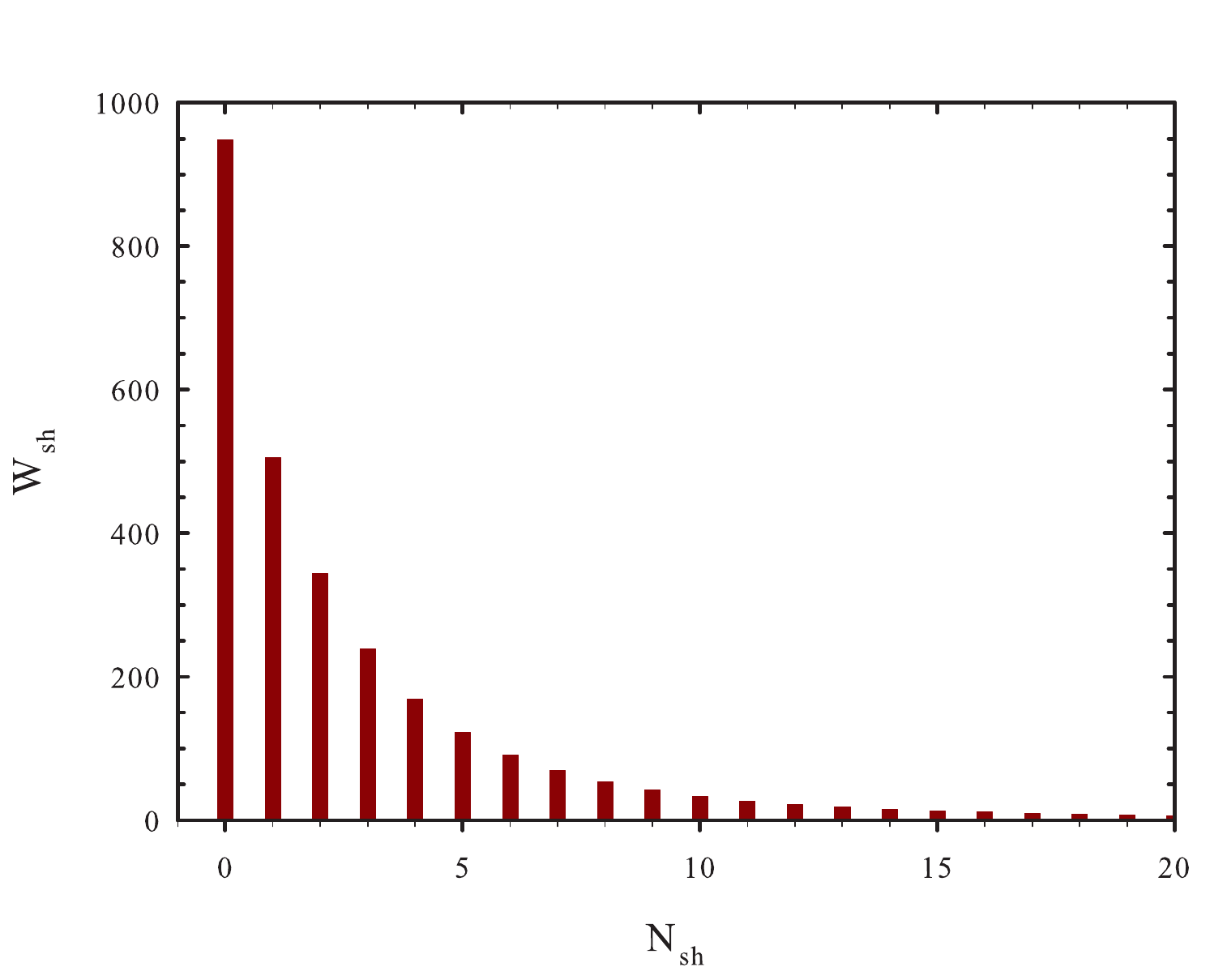}%
\caption{Structure of wave function of the $5/2^{-}$ resonance state in
$^{9}$B. Result is obtained with the MHNP.}%
\label{Fig:ShrellWeight52MRS9BMHN}%
\end{center}
\end{figure}

Let us consider candidates to the Hoyle analog states in $^{9}$B. This nuclei,
as was mentioned above, has no bound state and thus there is no way for
creating of a stable state in the triple collision of two alpha particles and
proton. However, one may consider the creation of a narrow resonance state in
$^{9}$B, which can be then transformed into a bound state of $^{9}$Be by 
emitting the
beta particle or in combination of  beta decay and gamma decay. This cascade of
decay can be considered as an additional alternatively way for synthesis of
$^{9}$Be nucleus. As we see from Table \ref{Tab:Resons9BExpMHNP}, there are two
very narrow resonance states in $^{9}$B, they are the 3/2$^{-}$ resonance
state with a ratio $\Gamma/E\approx$ 2.8$\cdot$10$^{-6}$ and the 5/2$^{-}$
resonance state having a ration $\Gamma/E\approx$ 6$\cdot$10$^{-3}$. The first
resonance state, as was shown in Figure \ref{Fig:ShrellWeight32MRSA9BHN}, is
very compact and very narrow three-cluster state. The 5/2$^-$ resonance state
in $^{9}$B is not so narrow as the 3/2$^{-}$ state, however, it has a wave
function with features which are typical for narrow resonance state (see
Figure \ref{Fig:ShrellWeight52MRS9BMHN}).%

Having calculated wave functions of a bound or resonance state, we can
evaluate the shape of a triangle, connected the centers of mass of interacting
clusters. In Table \ref{Tab:Size9Be9B} we display average distances
$R_{2}=R\left(  \alpha-\alpha\right)  $ and $R_{1}=R\left(  N-\alpha
\alpha\right)  $ between clusters. The quantity $R\left(  \alpha
-\alpha\right)  $ is an average distance between alpha particles, and
$R_{1}=R\left(  n-\alpha\alpha\right)  $ (or $R\left( p-\alpha\alpha\right) $)
determines the average distance between the neutron (proton)\ and the center
of mass of two alpha particles. It is important to note that to calculate
average distances for resonance states we make use the internal part of wave
functions, which we normalize to unity (see more detail in Refs.
\cite{2013UkrJPh.58.544V,2014PAN..77.555N} about definition of average
distances between clusters).%

\begin{table}[tbp] \centering
\caption{Avarage distances $R_1 =R(N-\alpha\alpha)$ and $R_2=R(\alpha-\alpha)$ 
 (in fm)  for few states of $^9Be$ and $^9B$.}%
\begin{tabular}
[c]{|c|c|c|c|c|}\hline
\multicolumn{5}{|c|}{$^{9}$Be}\\\hline
$J^{\pi}$ & $E$ & $\Gamma$ & $R_{1}$ & $R_{2}$\\\hline
$3/2^{-}$ & -1.574 &  & 3.71 & 3.38\\
$5/2^{-}$ & 0.897 & 2.363$\cdot$10$^{-5}$ & 4.75 & 3.52\\
$1/2^{+}$ & 0.338 & 0.168 & 14.02 & 7.37\\\hline
\multicolumn{5}{|c|}{$^{9}$B}\\\hline
$J^{\pi}$ & $E$ & $\Gamma$ & $R_{1}$ & $R_{2}$\\\hline
$3/2^{-}$ & 0.378 & 1.076$\cdot$10$^{-6}$ & 3.96 & 3.50\\\hline
$5/2^{-}$ & 2.805 & 0.017 & 5.01 & 3.87\\\hline
$1/2^{+}$ & 0.636 & 0.477 & 14.33 & 7.44\\\hline
\end{tabular}
\label{Tab:Size9Be9B}%
\end{table}%

Narrow resonance states, mentioned in Table \ref{Tab:Size9Be9B}, have
approximately the same size of a triangle as \ a bound state in $^{9}$Be.
Meanwhile, the 1/2$^{+}$ resonance states in $^{9}$Be and $^{9}$B are
represented by a very large triangle. We assume that such dispersed 
(even in the internal region) resonance states have very small 
probability (comparing to the very 
narrow resonance states) to be transformed into a bound state.

\section{Conclusions}

\label{Sect:Concl}

A three-cluster microscopic model was applied to studies of resonance 
states in
mirror nuclei $^{9}$Be and $^{9}$B. The model makes use of the hyperspherical
harmonics to numerate channels of three-cluster continuum and simplify the
method of solving of the Schr\"{o}dinger equation for a many-particle and
many-channel system. The MHNP modeled the
nucleon-nucleon interaction. It was shown that the model with such an NN
interaction provides a fairly good description of parameters of the known
resonance states. This potential provides much better description of the 
spectrum
of resonance states in $^{9}$Be and $^{9}$B, than the MP,
which was used in the previous paper \cite{2014PAN..77.555N}.  It was shown
that within the present model the excited 1/2$^{+}$ states in $^{9}$Be and
$^{9}$B\ are described as resonance states. It is also established that only 
one channel
with the hypermomentum $K=0$ dominates in formation and decay of these
$1/2^{+}$ resonance states. By analyzing effects of the Coulomb interaction,
we discovered three groups of resonance states which reveal weak, medium and
strong impact of the interaction on energy and width of resonance states. Our
analysis leads us to the conclusion that the very narrow 5/2$^{-}$ resonance
state in $^{9}$Be can be considered as the Hoyle-analogue state, and we assume
that this state is a key resonance state for the synthesis of $^{9}$Be in a
triple collision of alpha particles and neutron in a stellar environment.

\section*{Acknowledgment}

This work is partially supported by the Ministry of Education and Sciences of
Republic of Kazakhstan, the Research Grant IPS 3106/GF4.

\end{document}